\documentclass[]{article}
\usepackage{lmodern}
\usepackage{amssymb,amsmath}
\usepackage{ifxetex,ifluatex}
\usepackage{fixltx2e} 
\ifnum 0\ifxetex 1\fi\ifluatex 1\fi=0 
  \usepackage[T1]{fontenc}
  \usepackage[utf8]{inputenc}
\else 
  \ifxetex
    \usepackage{mathspec}
  \else
    \usepackage{fontspec}
  \fi
  \defaultfontfeatures{Ligatures=TeX,Scale=MatchLowercase}
\fi
\IfFileExists{upquote.sty}{\usepackage{upquote}}{}
\IfFileExists{microtype.sty}{%
\usepackage{microtype}
\UseMicrotypeSet[protrusion]{basicmath} 
}{}
\usepackage[margin=1in]{geometry}
\usepackage{hyperref}
\hypersetup{unicode=true,
            pdftitle={General Context-Aware Data Matching and Merging Framework},
            pdfauthor={Slavko Žitnik, Lovro Šubelj, Dejan Lavbič, Olegas Vasilecas and Marko Bajec},
            pdfborder={0 0 0},
            breaklinks=true}
\usepackage{natbib}
\usepackage{longtable,booktabs}
\usepackage{graphicx,grffile}
\makeatletter
\def\maxwidth{\ifdim\Gin@nat@width>\linewidth\linewidth\else\Gin@nat@width\fi}
\def\maxheight{\ifdim\Gin@nat@height>\textheight\textheight\else\Gin@nat@height\fi}
\makeatother
\setkeys{Gin}{width=\maxwidth,height=\maxheight,keepaspectratio}
\IfFileExists{parskip.sty}{%
\usepackage{parskip}
}{
\setlength{\parindent}{0pt}
\setlength{\parskip}{6pt plus 2pt minus 1pt}
}
\providecommand{\tightlist}{%
  \setlength{\itemsep}{0pt}\setlength{\parskip}{0pt}}
\ifx\paragraph\undefined\else
\let\oldparagraph\paragraph
\renewcommand{\paragraph}[1]{\oldparagraph{#1}\mbox{}}
\fi
\ifx\subparagraph\undefined\else
\let\oldsubparagraph\subparagraph
\renewcommand{\subparagraph}[1]{\oldsubparagraph{#1}\mbox{}}
\fi

\let\rmarkdownfootnote\footnote%
\def\footnote{\protect\rmarkdownfootnote}

\usepackage{titling}

  \title{General Context-Aware Data Matching and Merging Framework}
    \pretitle{\vspace{\droptitle}\centering\huge}
  \posttitle{\par}
    \author{Slavko Žitnik, Lovro Šubelj, Dejan Lavbič, Olegas Vasilecas and Marko
Bajec}
    \preauthor{\centering\large\emph}
  \postauthor{\par}
    \date{}
    \predate{}\postdate{}

\usepackage{booktabs}
\usepackage{longtable}
\usepackage{array}
\usepackage{multirow}
\usepackage[table]{xcolor}
\usepackage{wrapfig}
\usepackage{float}
\usepackage{colortbl}
\usepackage{pdflscape}
\usepackage{tabu}
\usepackage{threeparttable}
\usepackage{threeparttablex}
\usepackage[normalem]{ulem}
\usepackage{makecell}

\usepackage{amsthm}

\theoremstyle{definition}

\theoremstyle{definition}

\theoremstyle{definition}

\theoremstyle{remark}

\let\BeginKnitrBlock\begin \let\EndKnitrBlock\end
\begin{document}
\maketitle

\begin{quote}
Slavko Žitnik, Lovro Šubelj, \textbf{Dejan Lavbič}, Olegas Vasilecas and
Marko Bajec. 2013. \textbf{General Context-Aware Data Matching and
Merging Framework}, \href{https://www.mii.lt/informatica/}{Informatica
\textbf{(INFOR)}}, 24(1), pp.~119 - 152.
\end{quote}

\section*{Abstract}\label{abstract}
\addcontentsline{toc}{section}{Abstract}

Due to numerous public information sources and services, many methods to
combine heterogeneous data were proposed recently. However, general
end-to-end solutions are still rare, especially systems taking into
account different context dimensions. Therefore, the techniques often
prove insufficient or are limited to a certain domain. In this paper we
briefly review and rigorously evaluate a general framework for data
matching and merging. The framework employs collective entity resolution
and redundancy elimination using three dimensions of context types. In
order to achieve domain independent results, data is enriched with
semantics and trust. However, the main contribution of the paper is
evaluation on five public domain-incompatible datasets. Furthermore, we
introduce additional attribute, relationship, semantic and trust
metrics, which allow complete framework management. Besides overall
results improvement within the framework, metrics could be of
independent interest.

\section*{Keywords}\label{keywords}
\addcontentsline{toc}{section}{Keywords}

Entity resolution, redundancy elimination, semantic elevation, trust,
ontologies

\section{Introduction}\label{introduction}

Heterogeneous data matching and merging is due to increasing amount of
linked and open (on-line) data sources rapidly becoming a common need in
various fields. Different scenarios demand for analyzing heterogeneous
datasets collectively, enriching data with some on-line data source or
reducing redundancy among datasets by merging them into one. Literature
provides several state-of-the-art approaches for matching and merging,
although there is a lack of general solutions combining different
dimensions arising during the matching and merging execution. We propose
and evaluate a general and complete solution that allows a joint control
over these dimensions.

Data sources commonly include not only network data, but also data with
semantics. Thus a state-of-the-art solution should employ semantically
elevated algorithms (i.e.~algorithms that can process data with
semantics according to an ontology), to fully exploit the data at hand.
However, due to a vast diversity of data sources, an adequate data
architecture also employed. In particular, the architecture should
support all types and formats of data, and provide appropriate data for
each algorithm. As algorithms favor different representations and levels
of semantics behind the data, architecture should be structured
appropriately.

Due to different origin of (heterogeneous) data sources, the
trustworthiness (or accuracy) of their data can often be questionable.
Specially, when many such datasets are merged, the results are likely to
be inexact. A common approach for dealing with data sources that provide
untrustworthy or conflicting statements, is the use of trust management
systems and techniques. Thus matching and merging should be advanced to
a trust-aware level, to jointly optimize trustworthiness of data and
accuracy of matching or merging. Such collective optimization can
significantly improve over other approaches.

The paper proposes and demonstrates a general framework for matching and
merging execution. An adequate data architecture enables either pure
related data, in the form of networks, or data with semantics, in the
form of ontologies. Different datasets are merged using collective
entity resolution and redundancy elimination algorithms, enhanced with
trust management techniques. Algorithms are managed through the use of
different contexts that characterize each particular execution, and can
be used to jointly control various dimensions of variability of matching
and merging execution. Proposed framework is also fully implemented and
evaluated against real-world datasets.

The rest of the paper is structured as follows. The following section
gives a brief overview of the related work, focusing mainly on
trust-aware matching and merging. Next, section \ref{architecture},
presents employed data architecture and discusses semantic elevation of
the proposition. Section \ref{trust} formalizes the notion of trust and
introduces the proposed trust management techniques. General framework,
and accompanying algorithms, for matching and merging are presented in
section \ref{merging}. Experimental demonstration of the proposed
framework is shown in section \ref{experiments}, and further discussed
in section \ref{discussion}. Section \ref{conclusion} concludes the
paper.

\section{Related work}\label{related-work}

Recent literature proposes several state-of-the-art solutions for
matching and merging data sources. Analogous problems appear in many
different areas. When observing the area of matching and merging on a
broader basis, we used ideas from different approaches in the fields of
data integration
\citep{bhattacharya_iterative_2004, cohen_data_2000, hernandez_merge/purge_1995, lenzerini_data_2002},
data deduplication
\citep{ananthakrishna_eliminating_2002, kalashnikov_domain-independent_2006, monge_field_1996},
information retrieval, schema and ontology matching
\citep{castano_matching_2006, castano_dealing_2010, euzenat_ontology_2007, rahm_survey_2001},
and (related) entity resolution
\citep{bhattacharya_iterative_2004, bhattacharya_collective_2007}.

The propositions mainly address only selected issues of more general
matching and merging problem. In particular, approaches only partially
support the variability of the execution, commonly only homogeneous
sources, with predefined level of semantics, are employed, or the
approaches discard the trustworthiness of data and sources of origin. A
Mapping-based Object Matching - MOMA System \citep{thor_moma_2007}
presents the use of workflows and combination of several matching
algorithms within a single data source. Our approach uses attribute
resolution technique to align arbitrary data sources and prepares them
for further matching and merging techniques. The general problem of many
approaches over large-scale datasets is response time to first possible
results. Pay-As-You-Go ER \citep{whang_pay-as-you-go_2010} system
maximizes entity resolution progress with a limited amount of work
according to defined constraints. It orders merging pairs using these
constraints and outputs partial results as soon as possible. We run our
algorithms on network data and merge pairs according to similarity value
using contexts, where the user can observe the whole network during
merging and matching execution. Networks seemed the most appropriate to
design our approach. They enable us to dynamically change and read
structure as it is done by techniques of label propagation
\citep{subelj_robust_2011} or community detection
\citep{subelj_community_2011} where each community presents matched
data.

The proposed matching and merging approach employs the use of contexts
using semantics, trust and ontologies. The problem of matching
references to an underlying entity in natural language processing is
known as coreference resolution \citep{ng_unsupervised_2008}.
Traditionally the problem was solved using a set of constraints of
features, but improvements were achieved by using multiple matching
models and propagation of shared attributes across references
\citep{lee_stanfords_2011}. The idea of using different attribute,
related and semantic metrics is used from similar categorization of
features on simple pairwise approach which outperformed complex
coreference resolution models \citep{bengtson_understanding_2008}. Use
of ontologies, axioms and their inference as also used in text mining
\citep{stajner_entity_2009}, additionally gives us schema, knowledge
modelling and control mechanism \citep{lavbic_ontology-based_2010}
during matching and merging execution.

Literature also provides various trust-based, or trust-aware, approaches
for matching and merging
\citep{nagy_managing_2008, richardson_trust_2003}. Although they
formally exploit trust in the data, they do not represent a general or
complete solution. Mainly, they explore the idea of \emph{Web of Trust},
to model trust or belief in different entities. Related work on
\emph{Web of Trust} exists in the fields of identity verification
\citep{blaze_software_blaze_1999}, information retrieval
\citep{chakrabarti_automatic_1998}, social network analysis
\citep{domingos_mining_2001, kleinberg_authoritative_1999}, data mining
and pattern recognition
\citep{kautz_referral_1997, resnick_grouplens:_1994}. Our work also
relates to more general research of trust management and techniques that
provide formal means for computing with trust (e.g.
\citep{trcek_formal_2009}). Some research has also been done on using
the strategy of disinformation \citep{whang_disinformation_2013}. The
strategy focuses on matching and merging the records with bogus
information and is useful for robustness evaluation. The use of trust
management context in our approach is defined on levels from whole
source to attribute values.

This paper superseds our previously published theoretical concepts of
the same system \citep{subelj_merging_2011}. We did some minor changes
to definitions of ontology usage, renamed some notions (e.g.~Due to
disambiguation we are referring to relations as related data.) and
introduced an optimization by checking only neighbouring data
(\(19^{\text{th}}\) line of algorithm \ref{def:alg-er}). The main
contributions over the previous paper are experiments (see Section
\ref{experiments}) of all proposed methods and metrics on real datasets.
Implementations of general components are in-depth presented and
therefore it is shown the usage of semantics and trust improves overall
results.

\section{Data architecture}\label{architecture}

An adequate data architecture is of vital importance for efficient
matching and merging. Key issues arising are as follows:

\begin{enumerate}
\def\labelenumi{\arabic{enumi}.}
\tightlist
\item
  architecture should allow for data from heterogeneous sources,
  commonly in various formats,
\item
  semantical component of data should be addressed properly and
\item
  architecture should also deal with (partially) missing and uncertain
  data.
\end{enumerate}

To achieve superior performance, we propose a three level architecture
(see Figure \ref{fig:architecture}). Standard network data
representation on the bottom level (\emph{data level}) is enriched with
semantics (\emph{semantic level}) and thus elevated towards the topmost
real world level (\emph{abstract level}). Datasets on data level are
represented with networks, when the semantics are employed through the
use of ontologies.

Every dataset is (preferably) represented on data and semantic level.
Although both describe the same set of entities on abstract level, the
representation on each level is independent from others. This separation
resides from the fact that different algorithms of matching and merging
execution privilege different representations of data - either pure
related or semantically elevated representation. Separation thus results
in more accurate and efficient matching and merging, moreover,
representations can complement each other in order to boost the
performance.

The following section gives a brief introduction to networks, used for
data level representation. Section \ref{ontologies} describes ontologies
and semantic elevation of data level (i.e.~semantic level). Proposed
data architecture is formalized and further discussed in Section
\ref{levels}.

\subsection{Representation with
networks}\label{representation-with-networks}

\begin{figure}

{\centering \includegraphics[width=0.7\linewidth]{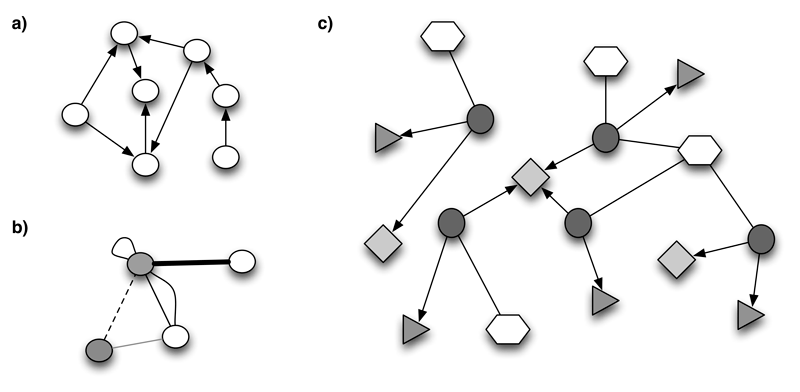}

}

\caption{(a) directed graph; (b) labeled undirected multigraph (labels are represented graphically); (c) network representing a group of related restaurants (circles correspond to restaurants, hexagons to their types, triangles to different phone numbers, while squares represent respective cities).}\label{fig:graphs}
\end{figure}

Most natural representation of any related data are \emph{networks}
\citep{newman_networks:_2010}. They are based upon mathematical objects
called \emph{graphs}. Informally speaking, graph consists of a
collection of points, called \emph{vertices}, and links between these
points, called \emph{edges} (see Figure \ref{fig:graphs}).

Let \(V_N\), \(E_N\) be a set of vertices, edges for some graph \(N\)
respectively. We define graph \(N\) as \(N=(V_N,E_N)\) where

\begin{equation}
V_N = \{v_1, v_2 \dots v_n\}
\label{eq:V-N}
\end{equation}

\begin{equation}
E_N \subseteq \{\{v_i, v_j\}|\mbox { } v_i, v_j \in V_N \wedge i<j \}
\label{eq:E-N}
\end{equation}

Edges are sets of vertices, hence they are not directed
(\emph{undirected graph}). In the case of \emph{directed graphs}
Equation \eqref{eq:E-N} rewrites to

\begin{equation}
E_N \subseteq \{(v_i, v_j)|\mbox { } v_i, v_j \in V_N \wedge i\neq j\}
\label{eq:E-N-direct}
\end{equation}

where \((v_i,v_j)\) is an edge from \(v_i\) to \(v_j\). The definition
can be further generalized by allowing multiple edges between two
vertices and \emph{loops} (edges that connect vertices with themselves).
Such graphs are called \emph{multigraphs} (see Figure \ref{fig:graphs}
b).

In practical applications we commonly strive to store some additional
information along with the vertices and edges. Formally, we define
\emph{labels} or \emph{weights} for each node and edge in the graph --
they represent a set of properties that can also be described using two
attribute functions

\begin{equation}
A_{V_N}: V_N\rightarrow\Sigma_1^{V_N}\times\Sigma_2^{V_N}\times\dots
\label{eq:A-V-N}
\end{equation}

\begin{equation}
A_{E_N}: E_N\rightarrow\Sigma_1^{E_N}\times\Sigma_2^{E_N}\times\dots
\label{eq:A-E-N}
\end{equation}

\(A_N=(A_{V_N}, A_{E_N})\), where \(\Sigma_i^{V_G}\), \(\Sigma_i^{E_G}\)
are sets of all possible vertex, edge attribute values respectively.

\emph{Networks} are most commonly seen as labeled, or weighted,
multigraphs with both directed and undirected edges (see Figure
\ref{fig:graphs} c). Vertices of a network represent some entities, and
edges represent related data between them. A (related) dataset,
represented with a network on the data level, is thus defined as
\((N,A_N)\).

\subsection{Semantic elevation using ontologies}\label{ontologies}

\begin{figure}

{\centering \includegraphics[width=0.8\linewidth]{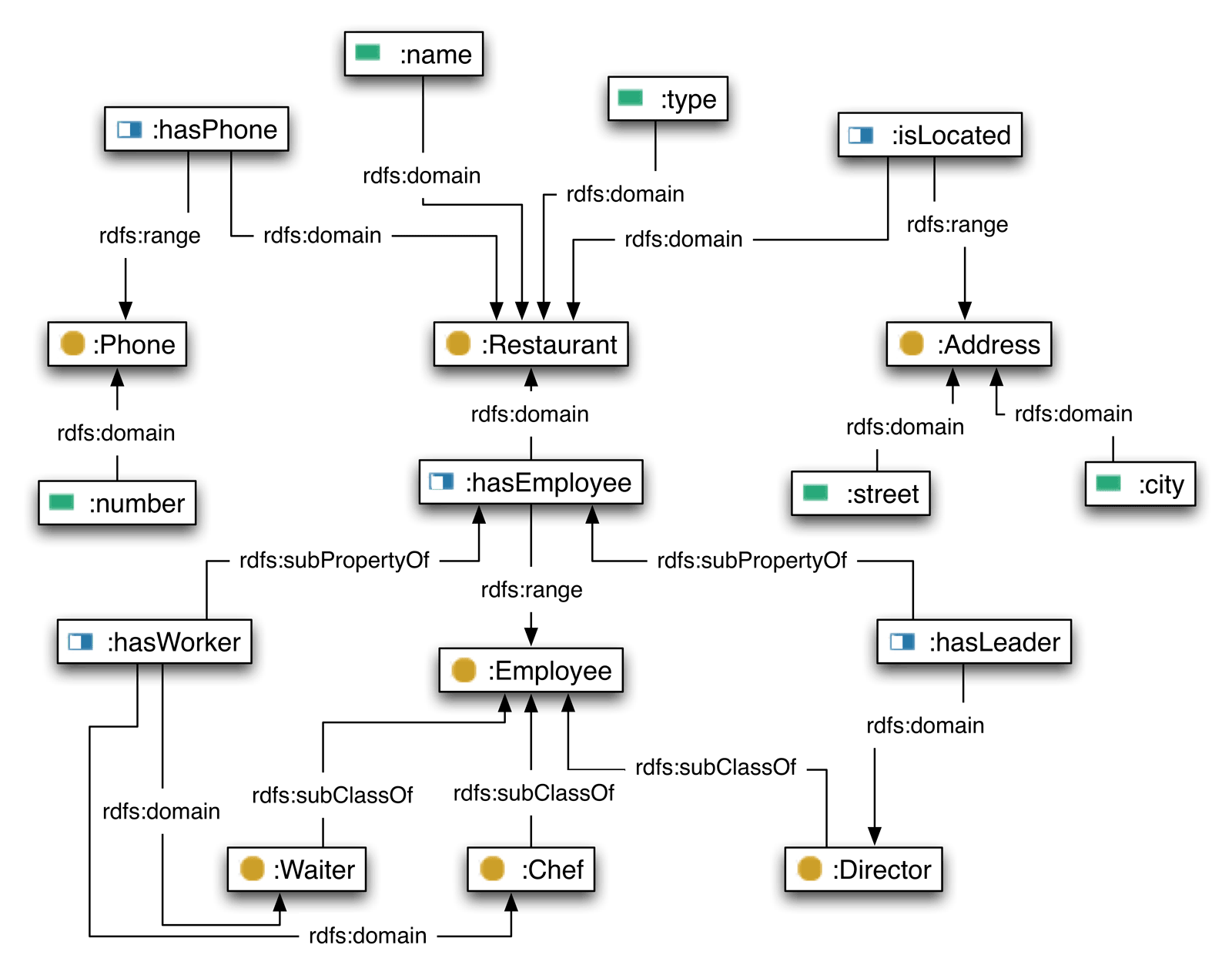}

}

\caption{A possible ontology over Restaurants dataset
(description in Section \ref{datasets}). Classes are represented by
circles, related data by half-white rectangles and attributes by
full-colour rectangles. Key concepts of the ontology are
\emph{Restaurant}, \emph{Address}, \emph{Phone} and \emph{Employee}.}\label{fig:ontology}
\end{figure}

\emph{Ontologies} are formal definitions of classes, related data,
functions and other objects. An \emph{ontology} is an explicit
specification of conceptualization \citep{gruber_translation_1993},
which is is an abstract view of the knowledge we wish to represent. It
can be defined as a network of \emph{entities}, restricted and annotated
with a set of \emph{axioms}. Let \(E_O\), \(A_O\) be the sets of
entities, axioms for some ontology \(O\) respectively. We propose a
dataset representation with an ontology on semantic level (an example in
Figure \ref{fig:ontology} as \(O=(E_O,A_O)\) where

\begin{equation}
E_O \subseteq E^C\cup E^I\cup E^R\cup E^A
\label{eq:E-O}
\end{equation}

\begin{equation}
A_O \subseteq \{a|\mbox{ }E_O^a\subseteq E_O\wedge a\mbox{ axiom on }E_O^a\}
\label{eq:A-O}
\end{equation}

Entities \(E_O\) consist of \emph{classes} \(E^C\) (concepts),
\emph{individuals} \(E^I\) (instances), \emph{related data} \(E^R\)
(among classes and individuals) and \emph{attributes} \(E^A\)
(properties of classes); and axioms \(A_O\) are assertions (over
entities) in a logical form that together comprise the overall theory
described by ontology \(O\).

This paper focuses on ontologies based on \emph{descriptive logic} that,
besides assigning meaning to axioms, enable also reasoning capabilities
\citep{horrocks_ontology_2001}. The latter can be used to compute
consequences of the previously made assumptions (queries), or to
discover non-intended consequences and inconsistencies within the
ontology.

One of the most prominent applications of ontologies is in the domain of
semantic interoperability (among heterogeneous software systems). While
pure semantics concerns the study of meanings, we define \emph{semantic
elevation} as a process to achieve semantic interoperability which be
considered as a subset of information integration.

Thus one of the key aspects of semantic elevation is to derive a common
representation of classes, individuals, related data and attributes
within some ontology. We employ a concept of \emph{knowledge chunks}
\citep{castano_dealing_2010}, where each entity is represented with its
\emph{name}, set of semantically \emph{related data}, attributes and
\emph{identifiers}. All of the data about a certain entity is thus
transformed into \emph{attribute-value} format, with an identifier of
the data source of origin appended to each value. Exact description of
the transformation between networked data and knowledge chunks is not
given, although it is very similar to the definition of inferred axioms
in Equation \eqref{eq:A-tilda-O}, section \ref{levels}. Knowledge chunks,
denoted \(k \in K\), thus provide a (common) synthetic representation of
an ontology that is used during the matching and merging execution. For
more details on knowledge chunks, and their construction from a
\emph{RDF(S)}\footnote{Resource Description Framework Schema} repository
or an \emph{OWL}\footnote{Web Ontology Language}, see
\citep{castano_dealing_2010, castano_icoord_2009}.

\subsection{Three level architecture}\label{levels}

\begin{figure}

{\centering \includegraphics[width=0.7\linewidth]{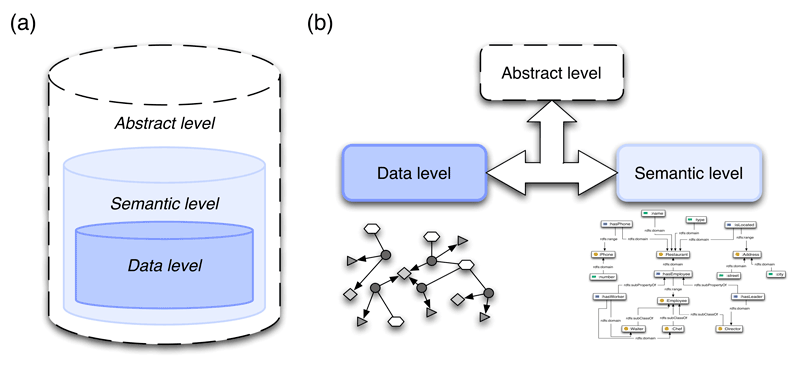}

}

\caption{(a) information-based view of the data architecture; (b) data-based view of the data architecture}\label{fig:architecture}
\end{figure}

As previously stated, every dataset is (independently) represented on
three levels -- \emph{data}, \emph{semantic} and \emph{abstract level}
(see Figure \ref{fig:architecture}). Bottommost \emph{data level} holds
data in a pure related format (i.e.~networks), mainly to facilitate
state-of-the-art related data algorithms for matching. Next level,
\emph{semantic level}, enriches data with semantics (i.e.~ontologies),
to further enhance matching and to promote semantic merging execution.
Data on both levels represent entities of topmost \emph{abstract level},
which serves merely as an abstract (artificial) representation of all
the entities, used during matching and merging execution.

The information captured by data level is a subset of that of semantic
level. Similarly, the information captured by semantic level is a subset
of that of abstract level. This \emph{information-based} view of the
architecture is seen in Figure \ref{fig:architecture} a). However,
representation on each level is completely independent from the others,
due to absolute separation of data. This provides an alternative
\emph{data-based} view, seen in Figure \ref{fig:architecture} b).

To manage data and semantic level independently (or jointly), a mapping
between the levels is required. In practice, data source could provide
datasets on both, data and semantic level. The mapping is in that case
trivial (i.e.~given). However, more commonly, data source would only
provide datasets on one of the levels, and the other has to be
\emph{inferred}.

Let \((N, A_N)\) be a dataset, represented as a network on data level.
Without loss for generality, we assume that \(N\) is an undirected
network. Inferred ontology
\((\tilde{E}_{\tilde{O}}, \tilde{A}_{\tilde{O}})\) on semantic level is
defined with

\begin{equation}
\tilde{E}_C = \{vertex, edge\}
\label{eq:E-C}
\end{equation}

\begin{equation}
\tilde{E}_I = V_N\cup E_N
\label{eq:E-I}
\end{equation}

\begin{equation}
\tilde{E}_R = \{isOf, isIn\}
\label{eq:E-R}
\end{equation}

\begin{equation}
\tilde{E}_A = \{A_{V_N}, A_{E_N}\}
\label{eq:E-A}
\end{equation}

and

\begin{equation}
\begin{split}
\tilde{A}_{\tilde{O}} = & \text{ } \{v \text{ isOf vertex } | \text{ } v \in V_N\} \\
 & \cup \text{ } \{e \text{ isOf edge } | \text{ } e \in E_N\} \\
 & \cup \text{ } \{v \text{ isIn } e \text{ } | \text{ } v \in V_N \wedge e \in E_N \wedge v \in e\} \\
 & \cup \text{ } \{v.A_{V_N} = a \text{ } | \text{ } v \in V_N\wedge A_{V_N}(v) = a\} \\
 & \cup \text{ } \{e.A_{E_N} = a \text{ } | \text{ } e \in E_N \wedge A_{E_N}(e) = a\}
\end{split}
\label{eq:A-tilda-O}
\end{equation}

We denote
\(\mathcal{I}_N: (N,A_N)\mapsto (\tilde{E}_{\tilde{O}},\tilde{A}_{\tilde{O}})\).
One can easily see that \(\mathcal{I}_N^{-1}\circ \mathcal{I}_N\) is an
identity (transformation preserves all the information).

On the other hand, given a dataset \((E_O,A_O)\), represented with an
ontology on semantic level, inferred (undirected) network
\((\tilde{N},\tilde{A}_{\tilde{N}})\) on data level is defined with

\begin{equation}
\tilde{V}_{\tilde{N}} = E_O\cap E^I
\label{eq:V-N-tilda}
\end{equation}

\begin{equation}
\tilde{E}_{\tilde{N}} = \{E_O^a\cap E^I|\mbox{ }a\in A_O\wedge E_O^a\subseteq E_O\}
\label{eq:E-N-tilda}
\end{equation}

and

\begin{equation}
\tilde{A}_{\tilde{V}_{\tilde{N}}}: \tilde{V}_{\tilde{N}}\rightarrow E^C\times E^A
\label{eq:A-V-N-tilda}
\end{equation}

\begin{equation}
\tilde{A}_{\tilde{E}_{\tilde{N}}}: \tilde{E}_{\tilde{N}}\rightarrow E^R
\label{eq:A-E-N-tilda}
\end{equation}

Instances of ontology are represented with the vertices of the network,
and axioms with its edges. Classes and related data are, together with
the attributes, expressed through vertex, edge attribute functions.

We denote
\(\mathcal{I}_O: (E_O,A_O)\mapsto (\tilde{N},\tilde{A}_{\tilde{N}})\).
Transformation \(\mathcal{I}_O\) discards purely semantic information
(e.g.~related data between classes), as it cannot be represented on the
data level. Thus \(\mathcal{I}_O\) cannot be inverted as
\(\mathcal{I}_N\). However, all the data, and data related information,
is preserved (e.g.~individuals, classes and related data among
individuals).

Due to limitations of networks, only axioms, relating at most two
individuals in \(E_O\), can be represented with the set of edges
\(\tilde{E}_{\tilde{N}}\) (see Equation \eqref{eq:E-N-tilda}). When this
is not sufficient, \emph{hypernetworks} (or \emph{hypergraphs}\footnote{Hypergraphs
  are graphs, where edges can connect multiple vertices.}) should be
employed instead. Nevertheless, networks should suffice in most cases.

One more issue has to be stressed. Although \(\mathcal{I}_N\) and
\(\mathcal{I}_O\) give a ``common'' representation of every dataset, the
transformations are completely different. For instance, presume
\((N,A_N)\) and \((E_O,A_O)\) are (given) representations of the same
dataset. Then \(\mathcal{I}_N(N,A_N)\neq (E_O,A_O)\) and
\(\mathcal{I}_O(E_O,A_O)\neq (N,A_N)\) in general -- inferred ontology,
network does not equal given ontology, network respectively. The former
non-equation resides in the fact that network \((N,A_N)\) contains no
knowledge of the (pure) semantics within ontology \((E_O,A_O)\); and the
latter resides in the fact that \(\mathcal{I}_O\) has no information of
the exact representation used for \((N,A_N)\). Still, transformations
\(\mathcal{I}_N\) and \(\mathcal{I}_O\) can be used to manage data on a
common basis.

Last, we discuss three key issues regarding an adequate data
architecture, presented in Section \ref{architecture}. Firstly, due to
variety of different data formats, a mutual representation must be
employed. As the data on both data and semantic level is represented in
the form of knowledge chunks (see Section \ref{ontologies}), every piece
of data is stored in exactly the same way. This allows for common
algorithms of matching and merging and makes the data easily manageable.

Furthermore, introduction of knowledge chunks naturally deals also with
missing data. As each chunk is actually a set of attribute-value pairs,
missing data only results in smaller chunks. Alternatively, missing data
could be randomly inputted from the rest and treated as extremely
uncertain or mistrustful (see Section \ref{trust}).

Secondly, semantical component of data should be addressed properly.
Proposed architecture allows simple (related) data and also semantically
enriched data. Hence no information is discarded. Moreover, appropriate
transformations make all data accessible on both data and semantic
level, providing for specific needs of each algorithm.

Thirdly, architecture should deal with (partially) missing and uncertain
or mistrustful data, which is thoroughly discussed in the following
section.

\section{Trust and trust management}\label{trust}

When merging data from different sources, these are often of different
origin and thus their trustworthiness (or accuracy) can be questionable.
For instance, personal data of participants in a traffic accident is
usually more accurate in the police record of the accident, then inside
participants' social network profiles. Nevertheless, an attribute from
less trusted data source can still be more accurate than an attribute
from more trusted one -- a related status (e.g.~single or married) in
the record may be outdated, while such type of information is inside the
social network profiles quite often up-to-date.

A complete solution for matching and merging execution should address
such problems as well. A common approach for dealing with data sources
that provide untrustworthy or conflicting statements, is the use of
\emph{trust management (systems)}. These are, alongside the concept of
trust, both further discussed in sections \ref{definition-trust} and
\ref{management-trust}.

\subsection{Definition of trust}\label{definition-trust}

\emph{Trust} is a complex psychological-sociological phenomenon. Despite
of, people use term trust in everyday life widely, and with very
different meanings. Most common definition states that trust is an
\emph{assured reliance on the character, ability, strength, or truth of
someone or something}.

In the context of computer networks, trust is modeled as a related data
between entities. Formally, we define a \emph{trust related data} as

\begin{equation}
\omega_E: E\times E\rightarrow \Sigma^E
\label{eq:omega-E}
\end{equation}

where \(E\) is a set of entities and \(\Sigma^E\) a set of all possible,
numerical or descriptive, trust values. \(\omega_E\) thus represents one
entity's attitude towards another and is used to model trust(worthiness)
of all entities in \(E\). To this end, different trust modeling
methodologies and systems can be employed, from qualitative to
quantitative (e.g.
\citep{nagy_managing_2008, richardson_trust_2003, trcek_formal_2009}).

We introduce trust on three different levels. First, we define trust on
the level of data source, in order to represent trustworthiness of the
source in general. Let \(S\) be the set of all data sources. Their trust
is defined as \(T_S: S\rightarrow [0,1]\), where higher values of
\(T_S\) represent more trustworthy source.

Second, we define trust on the level of attributes (or semantically
related data) within the knowledge chunks. The trust in attributes is
naturally dependent on the data source of origin, and is defined as
\(T_{A_s}: A_s\rightarrow [0,1]\), where \(A_s\) is the set of
attributes for data source \(s\in S\). As before, higher values of
\(T_{A_s}\) represent more trustworthy attribute.

Last, we define trust on the level of knowledge chunks. Despite the
trustworthiness of data source and attributes within some knowledge
chunk, its data can be (semantically) corrupted, missing or otherwise
unreliable. This information is captured using trustworthiness of
knowledge chunks, and again defined as \(T_{K}: K\rightarrow [0,1]\),
where \(K\) is a set of all knowledge chunks. Although the trust related
data (see Equation \eqref{eq:omega-E}), needed for the evaluation of
trustworthiness of data sources and attributes, are (mainly) defined by
the user, computation of trust in knowledge chunks can be fully
automated using proper evaluation function (see Section
\ref{management-trust}).

Three levels of trust provide high flexibility during matching and
merging. For instance, attributes from more trusted data sources are
generally favored over those from less trusted ones. However, by
properly assigning trust in attributes, certain attributes from else
less trusted data sources can prevail. Moreover, trust in knowledge
chunks can also assist in revealing corrupted, and thus questionable,
chunks that should be excluded from further execution.

Finally, we define trust in some particular value within a knowledge
chunk, denoted \emph{trust value} \(T\). This is the value in fact used
during merging and matching execution and is computed from corresponding
trusts on all three levels. In general, \(T\) can be an arbitrary
function of \(T_S\), \(T_{A_s}\) and \(T_K\). Assuming independence, we
calculate trust value by concatenating corresponding trusts

\begin{equation}
T = T_{S} \circ T_{A_s} \circ T_{K}
\label{eq:T}
\end{equation}

Concatenation function \(\circ\) could be a simple multiplication or
some fuzzy logic operation (trusts should in this case be defined as
fuzzy sets).

\subsection{Trust management}\label{management-trust}

During merging and matching execution, trust values are computed using
trust management algorithm based on \citep{richardson_trust_2003}. We
begin by assigning trust values \(T_S\), \(T_{A_s}\) for each data
source, attribute respectively (we actually assign trust related data).
Commonly, only a subset of values must necessarily be assigned, as
others can be inferred or estimated from the first. Next, trust values
for each knowledge chunk are not defined by the user, but are calculated
using the \emph{chunk evaluation function} \(f_{eval}\) (i.e.
\(T_K=f_{eval}\)).

An example of such function is a \emph{density of inconsistencies}
within some knowledge chunk. For instance, when attributes \emph{Birth}
and \emph{Age} of some particular knowledge chunk mismatch, this can be
seen as an inconsistency. However, one must also consider the trust of
the corresponding attributes (and data sources), as only inconsistencies
among trustworthy attributes should be considered. Formally, density of
inconsistencies is defined as

\begin{equation}
f_{eval}(k) = \frac{\hat{N}_{inc}(k)-N_{inc}(k)}{\hat{N}_{inc}(k)},
\label{eq:f-eval}
\end{equation}

where \(k\) is a knowledge chunk, \(k\in K\), \(N_{inc}(k)\) the number
of inconsistencies within \(k\) and \(\hat{N}_{inc}(k)\) the number of
all possible inconsistencies.

Finally, after all individual trusts \(T_S\), \(T_{A_s}\) and \(T_K\)
have been assigned, trust values \(T\) are computed using equation
\eqref{eq:T}. When merging takes place and two or more data sources (or
knowledge chunks) provide conflicting attribute values, corresponding to
the same (resolved) entity, trust values \(T\) are used to determine
actual attribute value in the resulting data source (or knowledge
chunk). For further discussion on trust management during matching and
merging see Section \ref{merging}.

\section{Matching and merging data sources}\label{merging}

Merging data from heterogeneous sources can be seen as a two-step
process. The first step resolves the real world entities of abstract
level, described by the data on lower levels, and constructs a mapping
between the levels. This mapping is used in the second step that
actually merges the datasets at hand. We denote these subsequent steps
as \emph{entity resolution} (i.e.~matching) and \emph{redundancy
elimination} (i.e.~merging).

Matching and merging is employed in various scenarios. As the specific
needs of each scenario vary, different dimensions of variability
characterize every matching and merging execution. These dimensions are
managed through the use of \emph{contexts}
\citep{castano_dealing_2010, lapouchnian_modeling_2009}. Contexts allow
a formal definition of specific needs arising in diverse scenarios and a
joint control over various dimensions of matching and merging execution.

The following section discusses the notion of contexts more thoroughly
and introduces different types of contexts used. Next, sections \ref{er}
and \ref{re} describe employed \emph{entity resolution} and
\emph{redundancy elimination} algorithms respectively. The general
framework for matching and merging is presented and formalized in
Section \ref{framework}, and discussed in Section \ref{discussion}.

\subsection{Contexts}\label{contexts}

\begin{figure}

{\centering \includegraphics[width=0.4\linewidth]{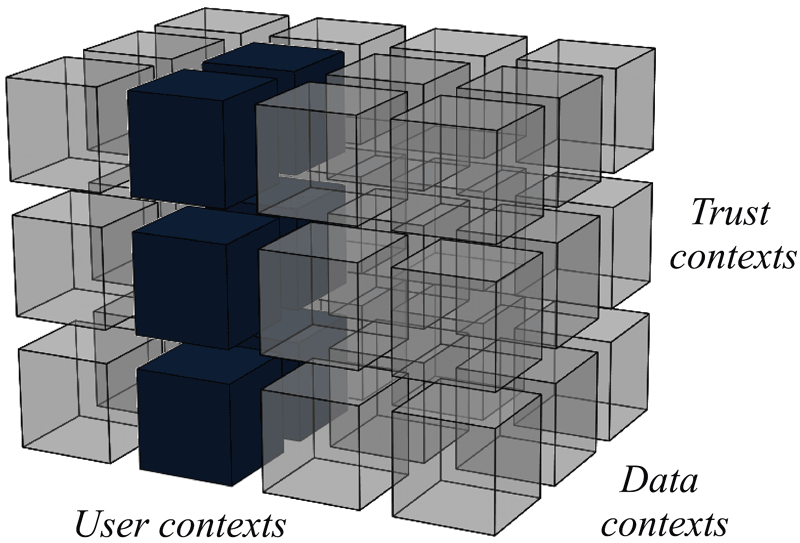}

}

\caption{Characterization of merging and matching execution defining one context in user dimension, two contexts in data dimension and all contexts in trust dimension}\label{fig:contexts}
\end{figure}

Every matching and merging execution is characterized by different
dimensions of variability of the data, and mappings between.
\emph{Contexts} are a formal representation of all possible operations
in these dimensions, providing for specific needs of each scenario.
Every execution is thus characterized with the contexts it defines (see
Figure \ref{fig:contexts}), and can be managed and controlled through
their use.

The idea of contexts originates in the field of requirements
engineering, where it has been applied to model domain variability
\citep{lapouchnian_modeling_2009}. It has just recently been proposed to
model also variability of the matching execution
\citep{castano_dealing_2010}. Our work goes one step further as it
introduces contexts, not bounded only to user or scenario specific
dimensions, but also data related and trust contexts.

Formally, we define a context \(C\) as

\begin{equation}
C: D \rightarrow \{true, false\},
\label{eq:C}
\end{equation}

where \(D\) can be any simple or composite domain. A context simply
limits all possible values, attributes, related data, knowledge chunks,
datasets, sources or other, that are considered in different parts of
matching and merging execution. Despite its simple definition, a context
can be a complex function. It is defined on any of the architecture
levels, preferably on all. Let \(C_A\), \(C_S\) and \(C_D\) represent
the same context on abstract, semantic and data level respectively. The
joint context is defined as

\begin{equation}
C_J = C_A \wedge C_S \wedge C_D
\label{eq:C-J}
\end{equation}

In the case of missing data (or contexts), only appropriate contexts are
considered. Alternatively, contexts could be defined as fuzzy sets, to
address also the noisiness of data. In that case, a fuzzy
\(\mathsf{AND}\) operation should be used to derive joint context
\(C_J\).

We distinguish between three types of contexts due to different
dimensions characterized (see Figure \ref{fig:contexts}).

\begin{itemize}
\tightlist
\item
  \textbf{User} or scenario specific contexts are used mainly to limit
  the data and control the execution. This type coincides with
  dimensions identified in \citep{castano_dealing_2010}. An example of
  user context is a simple selection or projection of the data.
\item
  \textbf{Data} related contexts arise from dealing with related or
  semantic data, and various formats of data. Missing or corrupted data
  can also be managed through the use of these contexts.
\item
  \textbf{Trust} and data uncertainty contexts provide for an adequate
  trust management and efficient security assurance between and during
  different phases of execution. An example of trust context is a
  definition of required level of trustworthiness of data or sources.
\end{itemize}

Detailed description of each context is out of scope of this paper. For
more details on (user) contexts see \citep{castano_dealing_2010}.

\subsection{Entity resolution}\label{er}

First step of matching and merging execution is to resolve the real
world entities on abstract level, described by the data on lower levels.
Thus a mapping between the levels (entities) is constructed and used in
consequent merging execution. Recent literature proposes several
state-of-the-art approaches for entity resolution (e.g.
\citep{ananthakrishna_eliminating_2002, bhattacharya_iterative_2004, bhattacharya_collective_2007, dong_reference_2005, kalashnikov_domain-independent_2006}.
A naive approach is a simple pairwise comparison of attribute values
among different entities. Although, such an approach could already be
sufficient for flat data, this is not the case for network data, as the
approach completely discards related data between the entities. For
instance, when two entities are related to similar entities, they are
more likely to represent the same entity. However, only the attributes
of the related entities are compared, thus the approach still discards
the information if related entities resolve to the same entities --
entities are even more likely to represent the same entities when their
related entities resolve to, not only similar, but the same entities. An
approach that uses this information, and thus resolves entities
altogether (in a collective fashion), is denoted \emph{collective}
(related) entity resolution algorithm.

We employ a state-of-the-art \emph{(collective) related data clustering}
algorithm proposed in \citep{bhattacharya_collective_2007}. To further
enhance the performance, algorithm is semantically elevated and adapted
to allow for proper and efficient trust management.

The algorithm \ref{def:alg-er} is actually a greedy agglomerative
clustering. Entities (on lower levels) are represented as a group of
clusters \(C\), where each cluster represents a set of entities that
resolve to the same entity on abstract level. At the beginning, each
(lower level) entity resides in a separate cluster. Then, at each step,
the algorithm merges two clusters in \(C\) that are most likely to
represent the same entity (most \emph{similar} clusters). When the
algorithm unfolds, \(C\) holds a mapping between the entities on each
level (i.e.~maps entities on lower levels through the entities on
abstract level).

During the algorithm, similarity of clusters is computed using a
\emph{joint similarity measure} (see Equation \eqref{eq:sim}), combining
\emph{attribute}, \emph{related data} and \emph{semantic similarity}.
First is a basic pairwise comparison of attribute values, second
introduces related information into the computation of similarity (in a
collective fashion), while third represents semantic elevation of the
algorithm.

Let \(c_i, c_j\in C\) be two clusters of entities. Using knowledge chunk
representation, attribute cluster similarity is defined as

\begin{equation}
sim_A(c_i, c_j) = \sum_{k_{i,j} \in c_{i,j} \wedge a \in k_{i,j}} trust(k_i.a, k_j.a) sim_A(k_i.a, k_j.a),
\label{eq:sim-A}
\end{equation}

where \(k_{i,j}\in K\) are knowledge chunks, \(a\in A_s\) is an
attribute and \(sim_A(k_i.a, k_j.a)\) similarity between two attribute
values. (Attribute) similarity between two clusters is thus defined as a
weighted sum of similarities between each pair of values in each
knowledge chunk. Weights are assigned due to trustworthiness of values
-- trust in values \(k_i.a\) and \(k_j.a\) is computed using

\begin{equation}
trust(k_i.a, k_j.a) = \min\{T(k_i.a), T(k_j.a)\}
\label{eq:trust}
\end{equation}

Hence, when even one of the values is uncertain or mistrustful,
similarity is penalized appropriately, to prevent matching based on
(likely) incorrect information.

For computation of similarity between actual attribute values
\(sim_A(k_i.a, k_j.a)\) (see Equation \eqref{eq:sim-A}), different
measures have been proposed. \emph{Levenshtein distance}
\citep{levenshtein_binary_1966} measures edit distance between two
strings -- number of insertions, deletions and replacements that
traverse one string into the other. Another class of similarity measures
are \emph{TF-IDF}\footnote{Term Frequency-Inverse Document Frequency}-based
measures (e.g. \emph{Cos TF-IDF} and \emph{Soft TF-IDF}
\citep{cohen_comparison_2003, moreau_robust_2008}). They treat attribute
values as a bag of words, thus the order of words in the attribute has
no impact on the similarity. Other attribute measures are also
\emph{Jaro} \citep{jaro_advances_1989} and \emph{Jaro-Winkler}
\citep{winkler_string_1990} that count number of matching characters
between the attributes.

Different similarity measures prefer different types of attributes.
\emph{TF-IDF}-based measures work best with longer strings
(e.g.~descriptions), when other prefer shorter strings (e.g.~names). For
numerical attributes, an alternative measure has to be employed
(e.g.~simple evaluation, followed by a numerical comparison). Therefore,
when computing attribute similarity for a pair of clusters, different
attribute measures are used with different attributes (see Equation
\eqref{eq:sim-A}).

Using data level representation, we define a \textit{neighborhood} for
vertex \(v\in V_N\) as

\begin{equation}
nbr(v) = \{v_n|\mbox{ }v_n\in V_N\wedge \{v, v_n\}\in E_N\}
\label{eq:nbr-v}
\end{equation}

and cluster \(c\in C\) as

\begin{equation}
nbr(c) = \{c_n|\mbox{ }c_n\in C\wedge v\in c\wedge c_n\cap nbr(v)\neq\emptyset\}.
\label{eq:nbr-c}
\end{equation}

Neighborhood of a vertex is defined as a set of connected vertices.
Similarly, neighborhood of a cluster is defined as a set of clusters,
connected through the vertices within.

For a (collective) related similarity measure, we adapt a \emph{Jaccard
coefficient} \citep{bhattacharya_collective_2007} measure for
trust-aware (related) data. \emph{Jaccard coefficient} is based on
\emph{Jaccard index} and measures the number of common neighbors of two
clusters, considering also the size of the clusters' neighborhoods --
when the size of neighborhoods is large, the probability of common
neighbors increases. We define

\begin{equation}
sim_R(c_i, c_j) = \frac{\sum_{c_n\in nbr(c_i)\cap nbr(c_j)} trust(e_{in}^T, e_{jn}^T)}{|nbr(c_i)\cup nbr(c_j)|}
\label{eq:sim-R}
\end{equation}

where \(e_{in}^T, e_{jn}^T\) is the most trustworthy edge connecting
vertices in \(c_n\) and \(c_i, c_j\) respectively (for the computation
of \(trust(e_{in}^T, e_{jn}^T)\), a knowledge chunk representation of
\(e_{in}^T, e_{jn}^T\) is used). (Related data) similarity between two
clusters is defined as the size of a common neighborhood (considering
also the trustworthiness of connecting related data), decreased due to
the size of clusters' neighborhoods. Entities related to a relatively
large set of entities that resolve to the same entities on abstract
level, are thus considered to be similar.

Alternatively, one could use some other similarity measure like
\emph{Adar-Adamic similarity} \citep{adamic_friends_2001}, random walk
measures, or measures considering also the ambiguity of attributes or
higher order neighborhoods \citep{bhattacharya_collective_2007}.

For the computation of the last, semantic, similarity, we propose a
random walk like approach. Using a semantic level representation of
clusters \(c_i,c_j\in C\), we do a number of random assumptions
(queries) over underlying ontologies. Let \(N_{ass}\) be the number of
times the consequences (results) of the assumptions made matched,
\(\tilde{N}_{ass}\) number of times the consequences were undefined (for
at least one ontology) and \(\hat{N}_{ass}\) the number of all
assumptions made. Furthermore, let \(N_{ass}^T\) be the trustworthiness
of ontology elements used for reasoning in assumptions that matched
(computed as a sum of products of trusts on the paths of reasoning,
similar as in Equation \eqref{eq:trust}). Semantic similarity is then
defined as

\begin{equation}
sim_S(c_i, c_j) = \frac{N_{ass}^T(c_i, c_j)}{\hat{N}_{ass}(c_i, c_j)-\tilde{N}_{ass}(c_i, c_j)}.
\label{eq:sim-S}
\end{equation}

Similarity represents the trust in the number of times ontologies
produced the same consequences, not considering assumptions that were
undefined for some ontology. As the expressiveness of different
ontologies vary, and some of them are even inferred from network data,
many of the assumptions could be undefined for some ontology. Still, for
\(\hat{N}_{ass}(c_i, c_j)-\tilde{N}_{ass}(c_i, c_j)\) large enough,
Equation \eqref{eq:sim-S} gives a good approximation of semantic
similarity.

Using attribute, related and semantic similarity (see Equations
\eqref{eq:sim-A}, \eqref{eq:sim-R} and \eqref{eq:sim-S}) we define a joint
similarity for two clusters as

\begin{equation}
\label{eq:sim}
sim(c_i, c_j) = \frac{\delta_A sim_A(c_i, c_j)+\delta_R sim_R(c_i, c_j)+\delta_S sim_S(c_i, c_j)}{\delta_A+\delta_R+\delta_S},
\end{equation}

where \(\delta_A\), \(\delta_R\) and \(\delta_S\) are weights, set due
to the scale of related and semantical information within the data. For
instance, setting \(\delta_R=\delta_S=0\) reduces the algorithm to a
naive pairwise comparison of attribute values, which should be used when
no related or semantic information is present.

\BeginKnitrBlock{definition}[Collective entity resolution]
\protect\hypertarget{def:alg-er}{}{\label{def:alg-er} {} }

\begin{align*}
&01: \quad \text{Initialize clusters as } C = \{\{k\} | \text{ } k \in K \} \\
&02: \quad \text{Initialize priority queue as } Q = \emptyset \\
&03: \quad \textbf{for } c_i, c_j \in C  \textbf{ and } sim(c_i, c_j) \ge \theta_S \\
&04: \quad \quad Q.insert(sim(c_i, c_j), c_i, c_j) \\
&05: \quad \textbf{end for} \\
&06: \quad \textbf{while } Q \neq \emptyset \textbf{ do} \\
&07: \quad \quad (sim(c_i, c_j), c_i, c_j) \leftarrow Q.pop() \quad \textit{// Most similar} \\
&08: \quad \quad \textbf{if } sim(c_i, c_j) < \theta_S \textbf{ then} \\
&09: \quad \qquad \textbf{return } C \\
&10: \quad \quad \textbf{end if} \\
&11: \quad \quad C \leftarrow C-\{c_i, c_j\} \cup \{c_i\cup c_j\} \quad \textit{// Matching} \\
&12: \quad \quad \textbf{for } (sim(c_x, c_k), c_x, c_k) \in Q \textbf{ and } x \in \{i,j\} \textbf{ do} \\
&13: \quad \qquad Q.remove(sim(c_x, c_k), c_x, c_k) \\
&14: \quad \quad \textbf{end for} \\
&15: \quad \quad \textbf{for } c_k \in C \textbf{ and } sim(c_i\cup c_j, c_k) \ge \theta_S \\
&16: \quad \qquad Q.insert(sim(c_i\cup c_j, c_k), c_i\cup c_j, c_k) \\
&17: \quad \quad \textbf{end for} \\
&18: \quad \quad \textbf{for } c_n \in nbr(c_i\cup c_j) \\
&19: \quad \qquad \textbf{for } c_k \in C \textbf{ and } sim(c_n, c_k) \ge \theta_S \\
&20: \quad \qquad \quad Q.insert(sim(c_n, c_k), c_n, c_k) \quad \textit{// Or update} \\
&21: \quad \qquad \textbf{end for} \\
&22: \quad \quad \textbf{end for} \\
&23: \quad \textbf{end while} \\
&24: \quad \textbf{return } C
\end{align*}
\EndKnitrBlock{definition}

Finally, we present the collective entity resolution algorithm
\ref{def:alg-er}. First, the algorithm initializes clusters \(C\) and
priority queue of similarities \(Q\), considering the current set of
clusters (lines \(1-5\)). Each cluster represents at most one entity as
it is composed out of a single knowledge chunk. Algorithm then, at each
iteration, retrieves currently the most similar clusters and merges them
(i.e.~matching of resolved entities), when their similarity is greater
than threshold \(\theta_S\) (lines \(7-11\)). As clusters are stored in
the form of knowledge chunks, matching in line 11 results in a simple
concatenation of chunks. Next, lines \(12-17\) update similarities in
the priority queue \(Q\), and lines \(18-22\) insert (or update) also
neighbors' similarities (required due to related similarity measure).
When the algorithm terminates, clusters \(C\) represent chunks of data
resolved to the same entity on abstract level. This mapping between the
entities (i.e.~their knowledge chunk representations) is used to merge
the data in the next step.

Threshold \(\theta_S\) represents minimum similarity for two clusters
that are considered to represent the same entities. Optimal value should
be estimated from the data.

Three more aspects of the algorithm ought to be discussed. Firstly,
pairwise comparison of all clusters during the execution of the
algorithm is computationally expensive, specially in early stages of the
algorithm. Authors in \citep{bhattacharya_collective_2007} propose an
approach in which they initially find groups of chunks that could
possibly resolve to the same entity. In this way, the number of
comparisons can be significantly decreased.

Secondly, due to the nature of (collective) related similarity measures,
they are ineffective when none of the entities has already been resolved
(e.g.~in early stages of the algorithm). As the measure in Equation
\eqref{eq:sim-R} counts the number of common neighbors, this always
evaluates to \(0\) in early stages (in general). Thus relative
similarity measures should be used after the algorithm has already
resolved some of the entities, using only attribute and semantic
similarities.

Thirdly, in the algorithm we implicitly assumed that all attributes,
(semantic) related data and other, have the same names or identifiers in
every dataset (or knowledge chunk). Although, we can probably assume
that all attributes within datasets, produced by the same source, have
same and unique names, this cannot be generalized.

We propose a simple, yet effective, solution. The problem at hand could
be denoted \emph{attribute resolution}, as we merely wish to map
attributes between the datasets. Thus we can use the approach proposed
for entity resolution. Entities are in this case attributes that are
compared due to their names, and also due to different values they hold;
and related data between entities (attributes) represent co-occurrence
in the knowledge chunks. As certain attributes commonly occur with some
other attributes, this would further improve the resolution.

Another possible improvement is to address also the attribute values in
a similar manner. As different values can represent the same underlying
value, \emph{value resolution}, done prior to attribute resolution, can
even further improve the performance.

\subsection{Redundancy elimination}\label{re}

After the entities, residing in the data, have been resolved (see
Section \ref{er}), the next step is to eliminate the redundancy and
merge the datasets at hand. This process is somewhat straightforward as
all data is represented in the form of knowledge chunks. Thus we merely
need to merge the knowledge chunks, resolved to the same entity on
abstract level. Redundancy elimination is done entirely on semantic
level, to preserve all the knowledge inside the data.

When knowledge chunks hold disjoint data (i.e.~attributes), they can
simply be concatenated together. However, commonly various chunks would
provide values for the same attribute and, when these values are
inconsistent, they need to be handled appropriately. A naive approach
would count only the number of occurrences of some value, when we
consider also their trustworthiness, to determine the most probable
value for each attribute.

\begin{figure}

{\centering \includegraphics[width=1\linewidth]{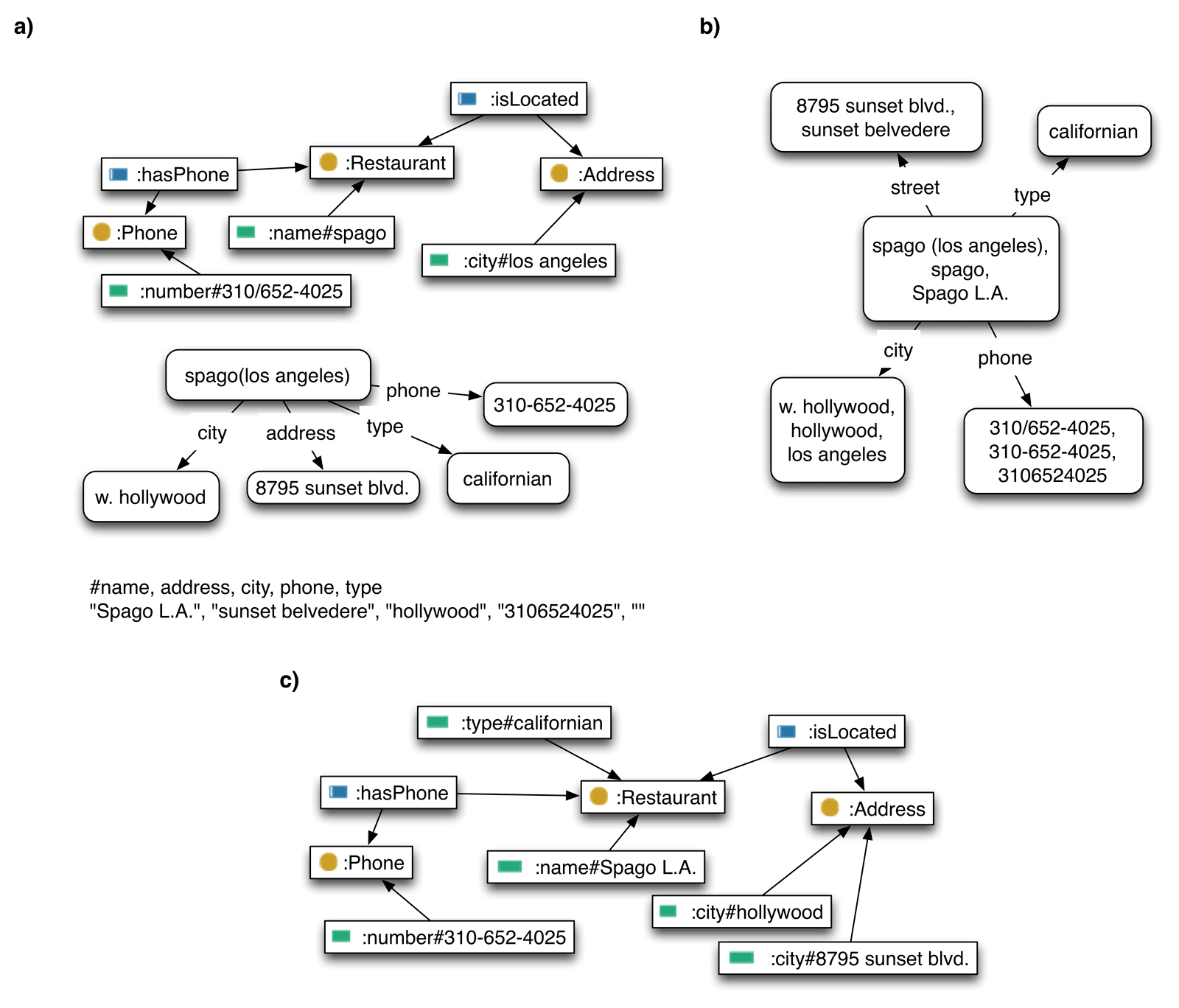}

}

\caption{Entity resolution and redundancy elimination on three
knowledge chunks (see Section \ref{ontologies}). a) Input data in a form
of ontology (see Figure \ref{fig:ontology}), network and attribute
values. b) Cluster network obtained with entity resolution
(i.e.~matching). c) Final ontology obtained after redundancy elimination
and appropriate postprocessing.}\label{fig:example}
\end{figure}

Let \(c\in C\) be a cluster representing some entity on abstract level
(resolved in the previous step), let \(k_1,k_2\dots k_n\in c\) be its
knowledge chunks and let \(k^c\) be the \emph{merged knowledge chunk},
we wish to obtain. Furthermore, for some attribute \(a \in A_ \cdot\),
let \(X^a\) be a random variable measuring the true value of \(a\) and
let \(X_i^a\) be the random variables for \(a\) in each knowledge chunk
it occurs (i.e. \(k_i.a\)). Value of attribute \(a\) for the merged
knowledge chunk \(k^c\) is then defined as

\begin{equation}
\operatorname*{arg\,max}_v P(X^a=v|\bigwedge_i X_i^a=k_i.a).
\label{eq:k-c-a-base}
\end{equation}

Each attribute is thus assigned the most probable value, given the
evidence observed (i.e.~values \(k_i.a\)). By assuming pair-wise
independence among \(X_i^a\) (conditional on \(X^a\)) and uniform
distribution of \(X^a\) equation \eqref{eq:k-c-a-base} simplifies to

\begin{equation}
\operatorname*{arg\,max}_v\prod_i P(X_i^a=k_i.a|X^a=v).
\label{eq:k-c-a-simp}
\end{equation}

Finally, conditional probabilities in equation \eqref{eq:k-c-a-simp} are
approximated with trustworthiness of values,

\begin{equation}
\label{eq:P-X-a}
P(X_i^a|X^a) \approx
\begin{cases}
T(k_i.a) & \text{for } k_i.a=v,\\
1-T(k_i.a) & \text{for } k_i.a\neq v
\end{cases}
\end{equation}

hence

\begin{equation}
k^c.a = \operatorname*{arg\,max}_v\prod_{k_i.a=v} T(k_i.a)\prod_{k_i.a\neq v} 1-T(k_i.a).
\label{eq:k-c-a}
\end{equation}

Only knowledge chunks (see section \ref{ontologies}) containing
attribute \(a\) are considered.

In the following we present the proposed redundancy elimination
algorithm \ref{def:alg-re}.

\BeginKnitrBlock{definition}[Redundancy elimination]
\protect\hypertarget{def:alg-re}{}{\label{def:alg-re} {} }

\begin{align*}
&1: \quad \text{Initialize knowledge chunks } K^C \\
&2: \quad \textbf{for } c \in C \textbf{ and } a \in A_\cdot \textbf{ do} \\
&3: \quad \quad k^c.a = \operatorname{arg\,max}_v\prod_{k\in c\wedge k.a=v} T(k.a)\prod_{k\in c\wedge k.a\neq v} 1-T(k.a) \\
&4: \quad \textbf{end for} \\
&5: \quad \textbf{return } K^C
\end{align*}
\EndKnitrBlock{definition}

The algorithm uses knowledge chunk representation of semantic level.
First, it initializes merged knowledge chunks \(k^c\in K^C\). Then, for
each attribute \(k^c.a\), it finds the most probable value among all
given knowledge chunks (line \(3\)). When the algorithm unfolds,
knowledge chunks \(K^C\) represent a merged dataset, with resolved
entities and eliminated redundancy. Each knowledge chunk \(k^c\)
corresponds to unique entity on abstract level, and each attribute holds
the most trustworthy value.

At the end, only the data that was actually provided by some data
source, should be preserved. Thus all inferred data (through
\(\mathcal{I}_N\) or \(\mathcal{I}_O\); see section \ref{levels}) is
discarded, as it is merely an artificial representation needed for
(common) entity resolution and redundancy elimination. Still, all
provided data and semantical information is preserved and properly
merged with the rest. Hence, although redundancy elimination is done on
semantic level, resulting dataset is given on both data and semantic
level (that complement each other).

Last, we discuss the assumptions of independence among \(X_i^a\) and
uniform distribution of \(X^a\). Clearly, both assumptions are violated,
still the former must be made in order for the computation of most
probable value to be feasible. However, the latter can be eliminated
when distribution of \(X^a\) can be approximated from some large-enough
dataset.

\subsection{General framework}\label{framework}

\begin{figure}

{\centering \includegraphics[width=1\linewidth]{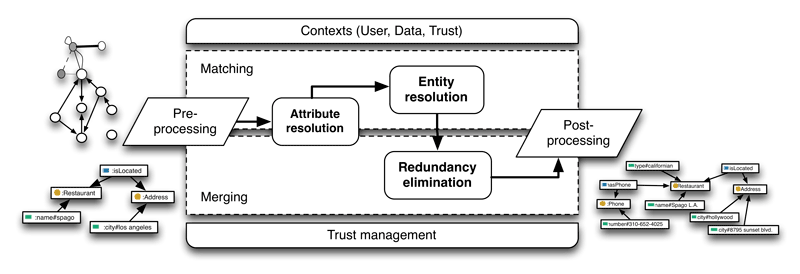}

}

\caption{General framework for matching and merging data from heterogeneous sources.}\label{fig:framework}
\end{figure}

Proposed entity resolution and redundancy elimination algorithms (see
sections \ref{er} and \ref{re}) are integrated into a general
\emph{framework} for matching and merging (see Figure
\ref{fig:framework}). Framework represents a complete solution, allowing
a joint control over various dimensions of matching and merging
execution. Each component of the framework is briefly presented in the
following, and further discussed in section \ref{discussion}.

Initially, data from various sources is preprocessed appropriately.
Every network or ontology is transformed into a knowledge chunk
representation and, when needed, also inferred on an absent architecture
level (see section \ref{levels}). After preprocessing is done, all data
is represented in the same, easily manageable, form, allowing for
common, semantically elevated, subsequent analyses.

Prior to entity resolution, attribute resolution is done (see section
\ref{er}). The process resolves and matches attributes in the
heterogeneous datasets, using the same algorithm as for entity
resolution. As all data is represented in the form of knowledge chunks,
this actually unifies all the underlying networks and ontologies.

Next, proposed entity resolution and redundancy elimination algorithms
are employed (see sections \ref{er} and \ref{re}). The process thus
first resolves entities in the data, and then uses this information to
eliminate the redundancy and to merge the datasets at hand. Algorithms
explore not only the related data, but also the semantics behind it, to
further improve the performance.

Last, postprocessing is done, in order to discard all artificially
inferred data and to translate knowledge chunks back to the original
network or ontology representation (see section \ref{architecture}).
Throughout the entire execution, components are jointly controlled
through (defined) user, data and trust contexts (see section
\ref{contexts}). Furthermore, contexts also manage the results of the
algorithms, to account for specific needs of each scenario.

Every component of the framework is further enhanced, to allow for
proper trust management, and thus also for efficient security assurance.
In particular, all the similarity measures for entity resolution are
trust-aware, moreover, trust is even used as a primary evidence in the
redundancy elimination algorithm. The introduction of trust-aware and
security-aware algorithms represents the main novelty of the
proposition.

\section{Experiments}\label{experiments}

In the following subsections we demonstrate the framework's (see Figure
\ref{fig:framework}) most important parts on several real-world
datasets, designed for entity resolution tasks and discuss the results.
The part of attribute resolution and redundancy elimination evaluation
is shown like a case study because to our knowledge, no tagged data
combining all results we need, exists.

The demonstration is done with respect to semantic elevation, semantic
similarity and trust management contexts (see section \ref{contexts}).
We do not fit methods for the datasets to achieve superior performance,
but rather focus on the increase of accuracy when using each of the
contexts. In the following we present the datasets, explain used
metrics, show the results and discuss them. The used datasets and full
source code is publicly available\footnote{\url{http://zitnik.si/mediawiki/index.php?title=Software}}.

\subsection{Datasets}\label{datasets}

We consider five testbeds of four different domains to simulate
real-life matching tasks. Each data source introduces many data quality
problems, in particular duplicate references, heterogeneous
representations, misspellings or extraction errors.

The CiteSeer dataset used is a cleaned version\footnote{\url{http://www.cs.umd.edu/projects/linqs/projects/er/DATA/citeseer.dat}}
from Getoor L. et. al. \citep{bhattacharya_collective_2007}, others were
presented and evaluated against entity resolution algorithms by
\citet{kopcke_evaluation_2010}\footnote{\url{http://dbs.uni-leipzig.de/en/research/projects/object_matching/fever/benchmark_datasets_for_entity_resolution}}.

\begin{itemize}
\tightlist
\item
  \textbf{CiteSeer} dataset contains \(1,504\) machine learning
  documents with \(2,892\) author references to \(1,165\) author
  entities. The only attribute information available is name for authors
  and title for documents.
\item
  \textbf{DBLP-ACM} dataset consists of two well-structured
  bibliographic data sources from DBLP and ACM with 2616 and 2294
  references to 2224 document entities. Each reference contains values
  for title, authors, venue and publication year of respective
  scientific paper.
\item
  \textbf{Restaurants} dataset contains \(864\) references to \(754\)
  restaurant entities. Most of the references contain values for name,
  address, city, phone number and type of certain restaurant.
\item
  \textbf{AbtBuy} is an e-commerce dataset with extracted data from
  \href{https://www.abt.com/}{Abt.com} and
  \href{https://www.buy.com}{Buy.com}. They contain \(1,081\) and
  \(1,092\) references to 1097 different products. Each product
  reference is mostly represented by product name, manufacturer and
  often missing description and price values.
\item
  \textbf{Affiliations} dataset consists of \(2,260\) references to
  \(331\) organizations. The only attribute value per record is an
  organization name, which can be written in many possible ways
  (i.e.~full, part name or abbreviation).
\end{itemize}

\subsection{Attribute resolution}\label{attr-res}

As a part of semantic elevation, the input datasets must be aligned by
\emph{attribute-value} pairs to achieve a mutual representation. As
mentioned in section \ref{er}, an entity resolution algorithm could be
used to merge appropriate attributes. To better solve the problem in
general, we propose the following similarity functions:

\begin{itemize}
\tightlist
\item
  \textbf{ExactMatch}: The simplest version. Attribute names must match
  exactly.
\item
  \textbf{SimilarityMatch}: Every two attributes with score above the
  selected threshold, are matched (We use \emph{Jaro-Winkler}
  \citep{winkler_string_1990} metric with threshold of \(0.95\)). This
  is typical pairwise entity resolution approach.
\item
  \textbf{SimilarityMatch+}: In addition to previous function, it
  considers synonyms when comparing two attribute names (synsets from
  semantic lexicon \emph{Wordnet} \citep{miller_wordnet:_1995} are
  used). Real-life datasets along with attribute names are created by
  people and that is why attributes over different datasets are supposed
  to be synonyms.
\item
  \textbf{DomainMatch}: Same attribute values contain similar data
  format. Leveraging this information we extract selected features and
  match the most similar attributes across datasets. (A simple example
  is calculating the average number of words per attribute values.)
\item
  \textbf{OntologyMatch}: Using ontologies, additional semantic
  information is included. If all input datasets are semantically
  described using ontologies, related data types \texttt{sameAs} or
  \texttt{seeAlso}, possible hierarchy of \emph{subclasses}, included
  rules and axioms can be additionally used for matching. When none of
  this apply, previous procedures must be employed.
\end{itemize}

As our datasets mostly consist of \(2\) different already aligned
sources, we have chosen some additional attribute names for
\emph{DBLP-ACM} dataset manually. Altered values are shown in Table
\ref{tab:ar}. Due to space limitations, we just presentively discuss the
results. First line are the original attribute names and next three
lines are changed to show success of proposed matchers. Pair \((1,2)\)
is successfully solved by \emph{SimilarityMatch}. The difference between
values is limited to misspellings and small writing errors. Pair
\((1,3)\) is a bit more difficult. Values \emph{authors} and
\emph{writers} or \emph{year} and \emph{yr} cannot be matched by
similarity. As they are synonyms, \emph{SimilarityMatch+} can match
them. Pair \((1,4)\) values are completely different and it is
completely useless to check name pairs. The \emph{DomainMatch} technique
correctly matches the attributes by considering attribute values format.

\begin{table}

\caption{\label{tab:ar}Attribute names, used to test attribute resolution
approaches on \emph{DBLP-ACM} dataset}
\centering
\begin{tabular}[t]{ccccc}
\toprule
\multicolumn{1}{c}{ } & \multicolumn{4}{c}{Attribute names} \\
\cmidrule(l{2pt}r{2pt}){2-5}
id & \#1 & \#2 & \#3 & \#4\\
\midrule
1 & title & venue & year & authors\\
2 & titl & venue & year & author\\
3 & title & venue & yr & writers\\
4 & attr1 & attr2 & attr3 & attr4\\
\bottomrule
\end{tabular}
\end{table}

\subsection{Entity resolution}\label{er-ex}

In this section we first discuss the selected entity resolution
algorithm and then show the increase in correctly matched values using
semantic similarity. We implemented the algorithm, proposed in
\citep{bhattacharya_collective_2007}, which is well described in section
\ref{er} and presented as algorithm \ref{def:alg-re}.

In addition to the standard blocking techniques of partially string
matching we added similarity, n-gram blocking and also enabled the
option of fuzzy blocking. Standard approach is used on \emph{AbtBuy},
\emph{CiteSeer} and \emph{DBLP-ACM} datasets. Similarity blocking adds
an instance to a block if the similarity score with the representative
reference of the block is above the defined threshold. This type of
blocking was used with the \emph{Restaurants} dataset using \(0.3\)
threshold for \emph{name} and \(0.7\) for \emph{phone} attribute. At
\emph{Affiliations} dataset we use n-gram blocking with at least \(4\)
\(6\)-gram matches.

We use secondstring \citep{cohen_comparison_2003} library for all basic
similarity measures implementations. At bootstrapping and clustering we
use JaroWinkler with TFIDF and manual weights as an attribute metric.
Promising general results were achieved also using n-gram and
Level2JaroWinkler metric. As a related similarity we use k-Neighbours at
bootstrapping and modified JaccardCoefficient at clustering. The
modification just aligns the match result \(x\) using function
\(f(x) = -(-x+1)^{10}+1\), because similarity pairs instead of typical
sets are checked.

\begin{figure}

{\centering \includegraphics[width=0.8\linewidth]{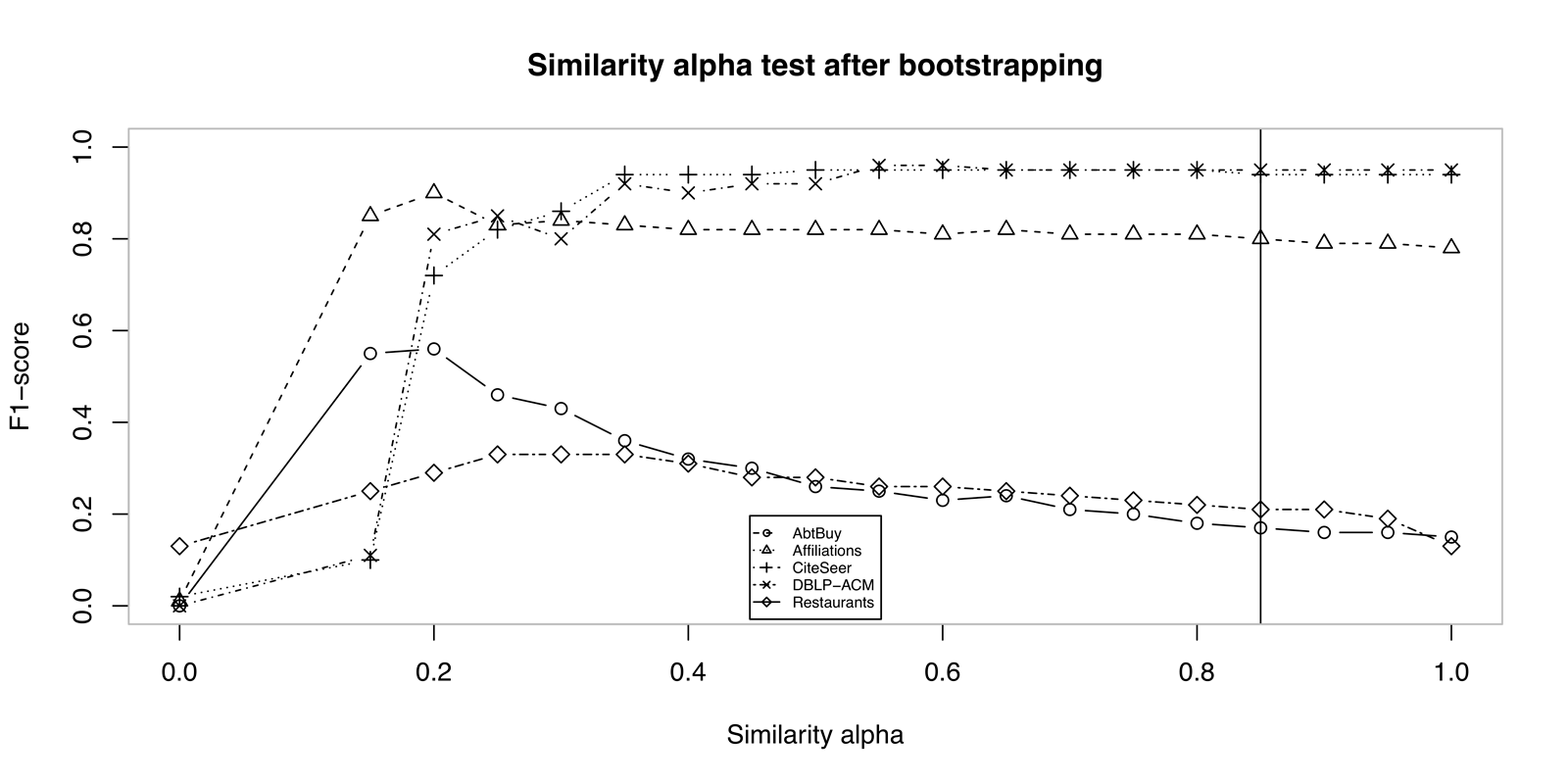}

}

\caption{Comparison of entity resolution results after bootstrapping according to similarity alpha.}\label{fig:erAlphClus}
\end{figure}

The most important parameters that need to be selected are similarity
alpha \(\alpha\) and merge threshold \(\theta_S\). Both values were
selected subjectively and not dataset - specific. We set similarity
alpha to \(\alpha = 0.85\), which results in weighting attribute metric
to \(\delta_A = \alpha\) and related data metric to
\(\delta_R = 1-\alpha\). Matching accuracy using different similarity
alphas is shown on Figure \ref{fig:erAlphClus}. As it can be seen from
the figure, some datasets contain a lot of disambiguate values, which
results in very low F-score at \(\alpha\) set to \(1\).

\begin{figure}

{\centering \includegraphics[width=0.8\linewidth]{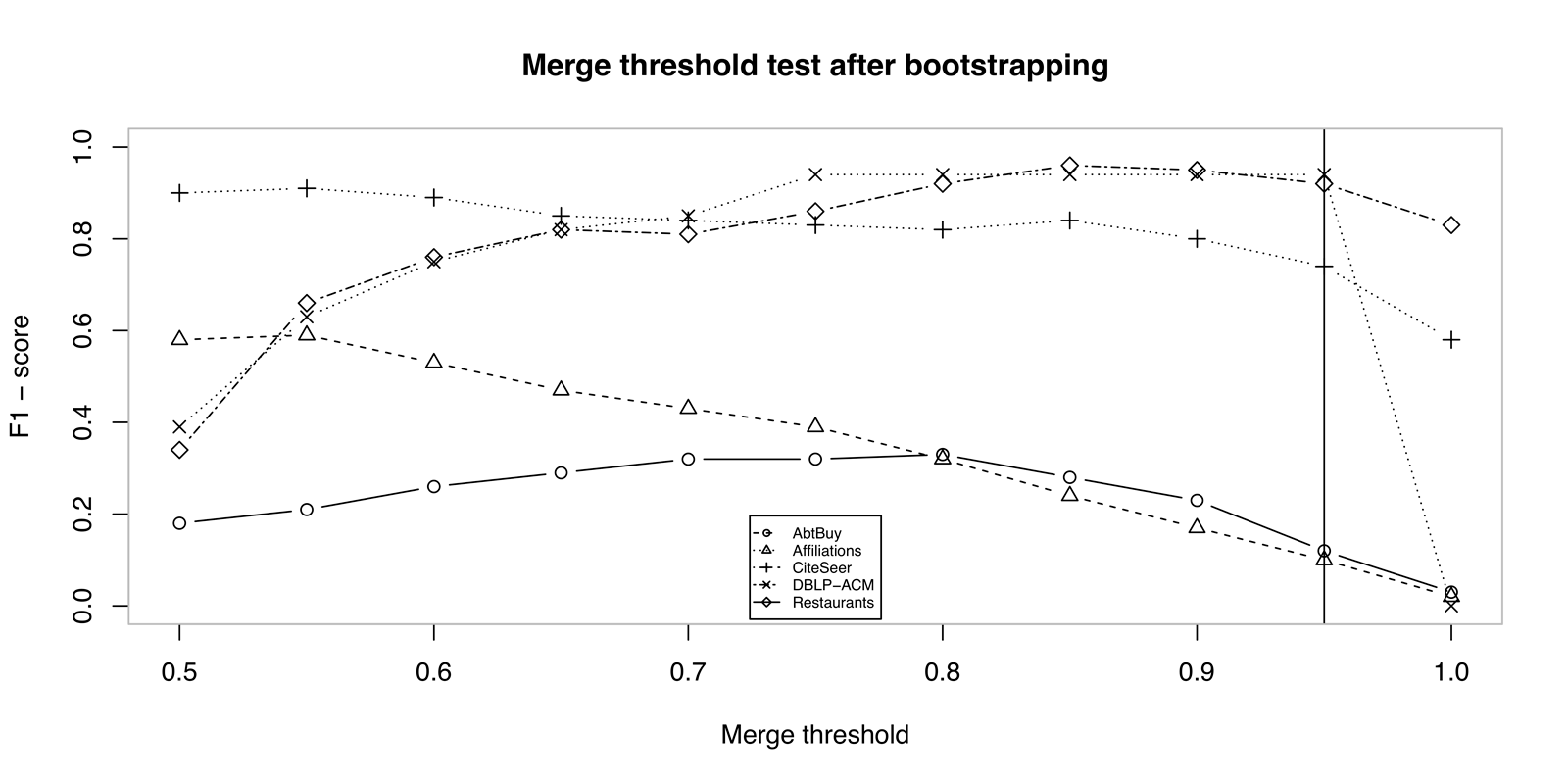}

}

\caption{Comparison of entity resolution results after bootstrapping according to $\theta_S$ merge threshold.}\label{fig:erThreBoot}
\end{figure}

\begin{figure}

{\centering \includegraphics[width=0.8\linewidth]{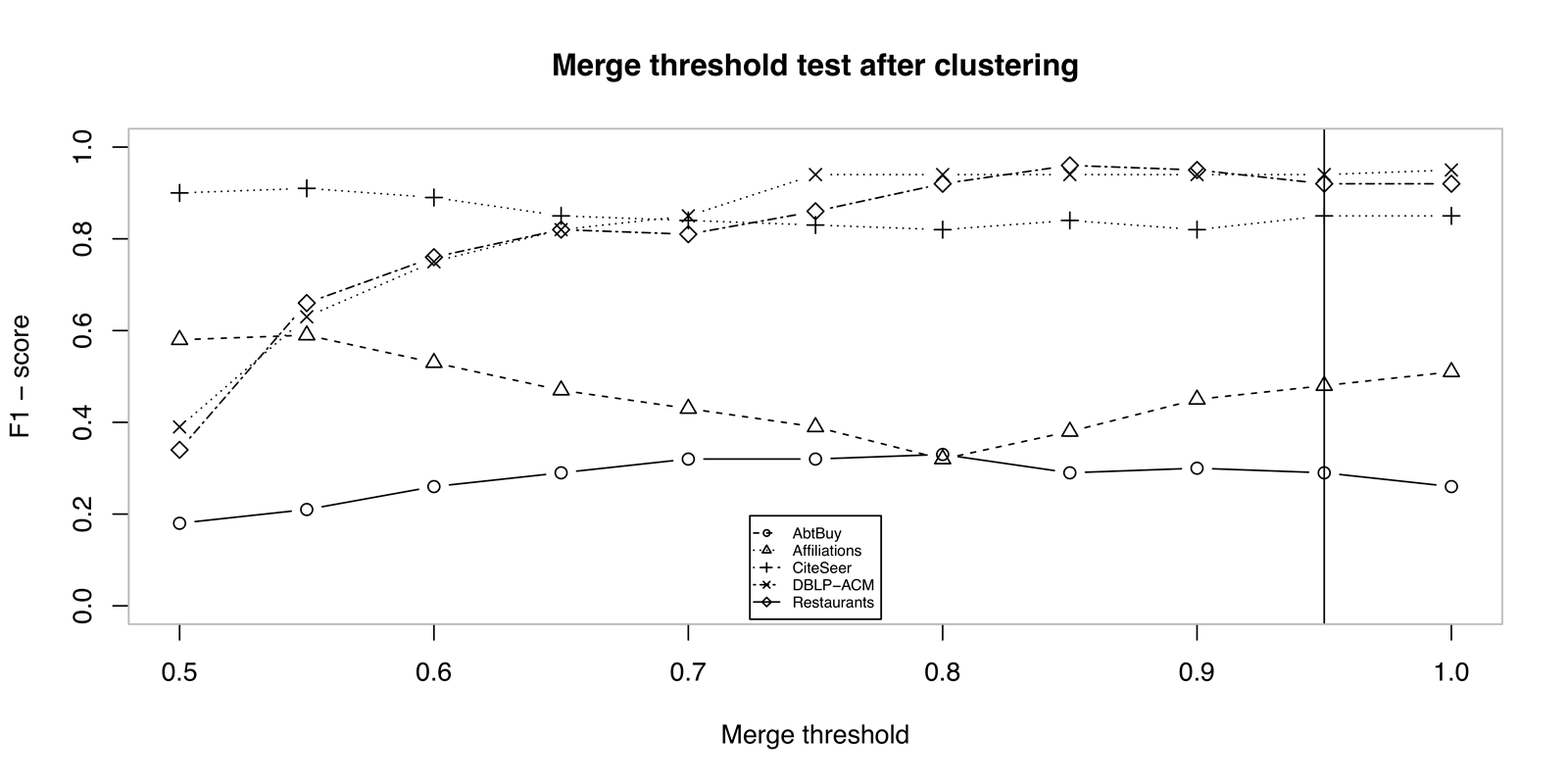}

}

\caption{Comparison of entity resolution results after clustering according to $\theta_S$ merge threshold.}\label{fig:erThreClus}
\end{figure}

Merge threshold in our solution is set to \(0.95\). Testing the
threshold at different values after bootstrapping is presented on Figure
\ref{fig:erThreBoot} and after clustering on Figure
\ref{fig:erThreClus}. It is possible to see the effect of iterative
matching and related data metric from the Figure \ref{fig:erThreClus},
which improves the final results. During testing these parameters, no
semantic measure was used yet.

Due to optimization, our implementation updates or possibly inserts only
neighbour pairs of matched clusters into priority queue during
clustering. The accuracy when checking only neighbours remains
unchanged. Therefore the cluster \(c_k \in nbr(c_i \cup c_j)\) at the
\(19^{th}\) line of algorithm \ref{def:alg-er}).

\begin{figure}

{\centering \includegraphics[width=1\linewidth]{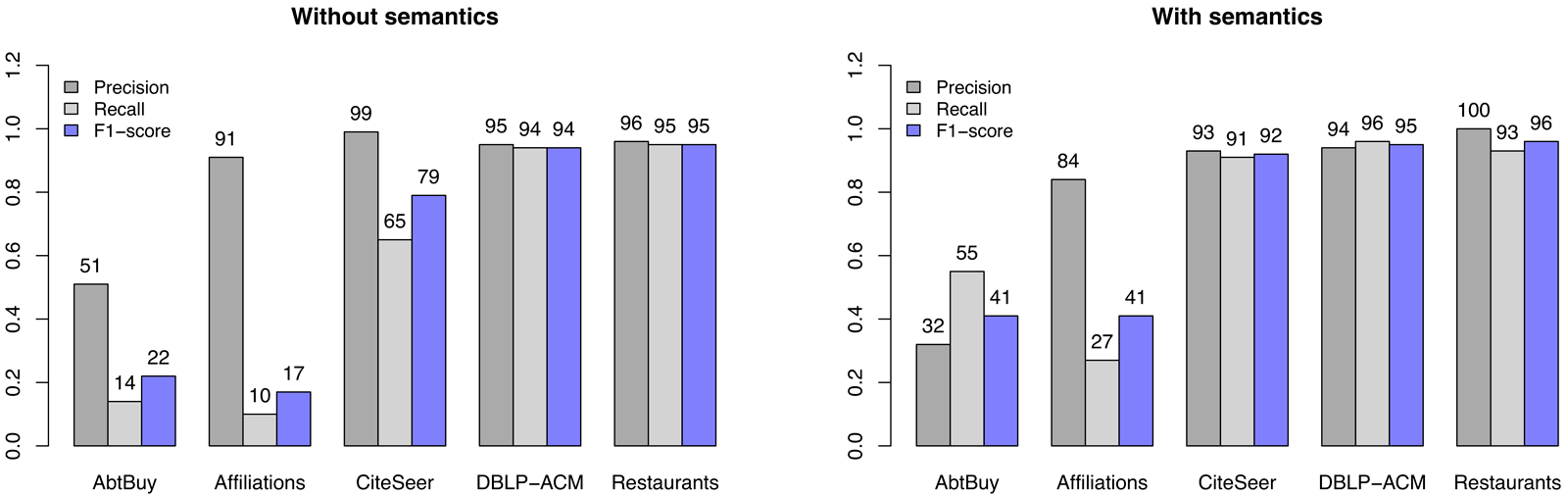}

}

\caption{Comparison of entity resolution results after bootstrapping without and with using semantic similarity.}\label{fig:erCompBoot}
\end{figure}

\begin{figure}

{\centering \includegraphics[width=1\linewidth]{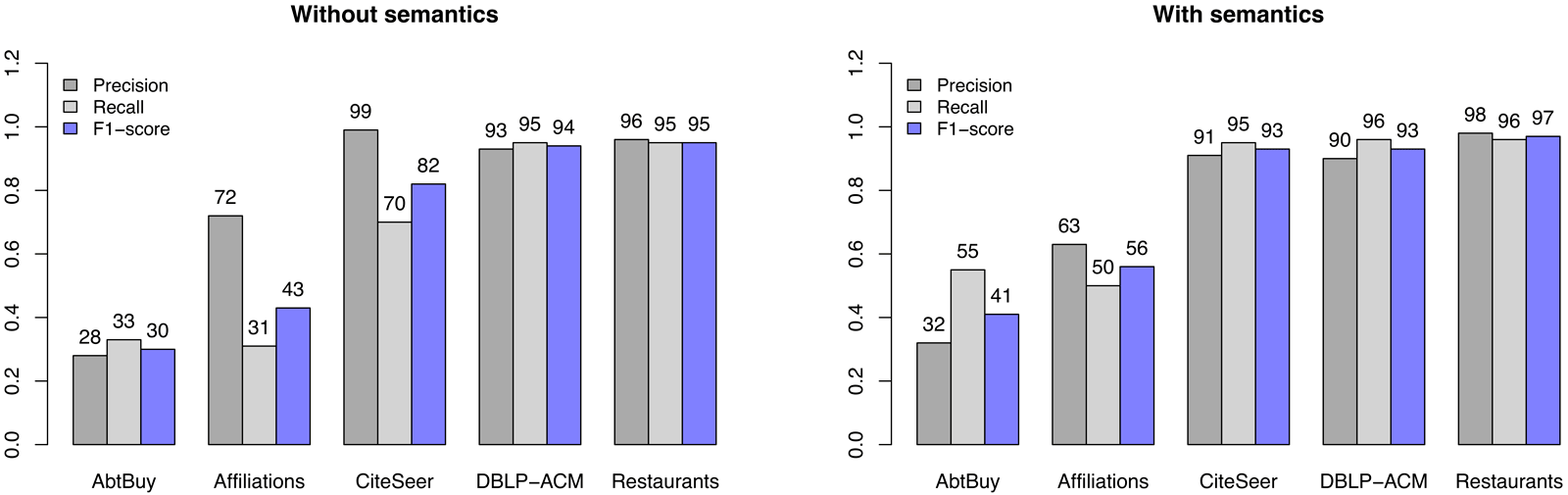}

}

\caption{Comparison of entity resolution results after clustering without and with using semantic similarity.}\label{fig:erCompClus}
\end{figure}

On Figures \ref{fig:erCompBoot} and \ref{fig:erCompClus} we present the
increase of success in matching using semantic similarity (see Equation
\eqref{eq:sim-S}). We set semantic similarity weight \(\delta_s\) to \(4\)
based on some preliminary experiments. Getoor et. al.
\citep{bhattacharya_collective_2007} adjusted Adar
\citep{adamic_friends_2001} similarity metric to better support values
(e.g.~author names) disambiguation. It learns an ambiguity function
after checking the whole set of values in the dataset, similar to TF-IDF
approach. This metric better models names, but does not use semantics,
like identifying first or last name, product codes or specific parts of
given value. Semantic similarity should model the human reasoning
whether to match two values or not. For better understanding the meaning
of semantic similarity, we present few examples, used in the experiment:

\begin{itemize}
\tightlist
\item
  \textbf{Name metric}: This is our the most general similarity metric.
  It models typical value matching by splitting it into tokens,
  identifying the value with less information and comparing it to other
  value's tokens by \emph{startWith} or similarity metric. It also
  checks and matches abbreviations. For example, every pair of names
  ``William Cohen'', ``W. Cohen'', ``W.W. Cohen'' or ``Cohen'' must have
  maximum semantic similarity. Similar applies to ``Arizona State Univ.,
  Tempe, AZ'', ``Arizona State University'' and ``Arizona State
  University, Arizona'' where using string similarity metric yields low
  values. Name metric is used for organization name matching on
  \emph{Affilation} dataset, author name matching on \emph{CiteSeer} and
  phone and restaurant name matching on \emph{Restaurants} dataset.
\item
  \textbf{Number metric}: Number metric identifies numeric values and
  matches them according to difference in values. It is used on
  \emph{DBLP-ACM} dataset at publication year matching.
\item
  \textbf{Product metric}: Product metric is designed to match products,
  which sometimes contain serial numbers or codes. These codes are
  commonly represented as a sequence of numbers and/or letters. In
  addition to code matching, it integrates Name metric with minimum
  \(k\) token match score. An example of matching two products is
  ``Toshiba 40' Black Flat Panel LCD HDTV - 40RV525U'' and ``Toshiba
  40RV525U - 40' Widescreen 1080p LCD HDTV w/ Cinespeed - Piano Black''
  where it is very hard to identity pair without code detection. Product
  metric is used on \emph{AbtBuy} dataset for product name matching.
\item
  \textbf{Restaurant metric}: This metric is specific to
  \emph{Restaurants} dataset. It supposes attributes name and phone or
  location are scored above the threshold to match.
\item
  \textbf{Title metric}: Titles are sometimes shrinked, have some words
  replaced with synonyms or refer to papers, written in more parts. This
  metric improves matching titles on \emph{DBLP-ACM dataset}.
\end{itemize}

The results on Figures \ref{fig:erCompBoot} and \ref{fig:erCompClus}
show the increase of matching accuracy by employing semantic similarity
measure. The results on \emph{AbtBuy} dataset are increased by \(11\%\)
after clustering. Recall is significantly higher, but precision falls
down. It is interesting that using semantics, same result is achieved
immediately after bootstrapping, which shows good work at blocking. On
\emph{Affiliations}, the precision lowers, but employing semantics, more
organizations with different name representations are resolved.
\emph{CiteSeer} gains more than \(10\%\) in recall and F-score and also
keeps all measures above \(90\%\). At \emph{DBLP-ACM} dataset, the
differences are not very significant, but use of semantics still shows
minor improvements. After bootstrapping at \emph{CiteSeer} dataset it is
interesting semantic similarity helps achieving \(100\%\) precision and
a little improves the final result.

Experimenting using only semantic similarity metric gave worse results
than including also attribute one. This is because our semantic
similarities focus on semantics and not on misspelled or disambiguated
data on lexical level. \emph{Restaurant} dataset for example contains
examples unsolvable even for a man without background knowledge. In the
case of \emph{AbtBuy} dataset even more knowledge would not work as name
and product description is too general to match on some examples. Number
match metric could be applied also on it, but one of the datasets barely
contains a product's price.

\subsection{Redundancy elimination}\label{re-ex}

The last step before postprocessing is merging knowledge chunks matched
in clusters at entity resolution.

Merging is done entirely using trust management. In section \ref{trust}
we define trust on levels of data source, knowledge chunk and value. As
trust cannot be easily initialized, we select the appropriate cluster
representative using trust of value only. Therefore we implemented the
calculation of trust value for algorithm \ref{def:alg-re} in the
following ways:

\begin{itemize}
\tightlist
\item
  \textbf{Random}: Random value is selected as the representative.
\item
  \textbf{Naive}: Value that occurs the most time is selected as the
  representative.
\item
  \textbf{Naive+}: The representative is selected as the maximum similar
  value to all others. Let \(c\) be a cluster of matched values, \(k\)
  value in cluster and Sim appropriate similarity function. Then the
  value is selected according to Equation \eqref{eq:naivep}. As similarity
  function we use \emph{Jaro-Winkler}.
\end{itemize}

\begin{equation}
{\rm Representative}(c) = \operatorname{arg\,max}_v \sum_{k \in c \wedge k \neq v} {\rm Sim}(k,v)
\label{eq:naivep}
\end{equation}

\begin{itemize}
\tightlist
\item
  \textbf{Trust}: Intuitively, a value is trustworthy if it yields many
  search results on the internet. This is not exactly true as for
  example the number of search results for ``A. N.'' is much higher
  comparing to ``Andrew Ng''. By investigating some person name - based
  test searches, we expect the number of search hits decreases a lot if
  the word is misspelled. We denote \(N_{hits}(v)\) as the number of
  hits for value \(v\). Let \(N_{nhits}(v)\) be number of hits for a
  value of \(v\) with some noise added. We set \(m\) to \(5\) and change
  \(4\) letters or numbers randomly. The trust is calculated as in
  Equation \eqref{eq:trustre} and as the result, the maximum trust value
  is selected.
\end{itemize}

\begin{equation}
{\rm Trust}(v) = 1-\frac{\sum_{1 \le i \le m}N_{nhits}(v)}{m \cdot N_{hits}(v)}
\label{eq:trustre}
\end{equation}

During experiments, random clusters, having more than \(10\) values of
specific attribute were selected for redundancy elimination. Using
clusters with multiple values, the results are more representative
because it is harder to select the right value. In each cluster, we add
noise to a portion of values. So, one of the non-noise values is
expected to be returned as a result of redundancy elimination because
they certainly better represent the entity and this is taken as a
measure of classification accuracy.

\begin{figure}

{\centering \includegraphics[width=0.8\linewidth]{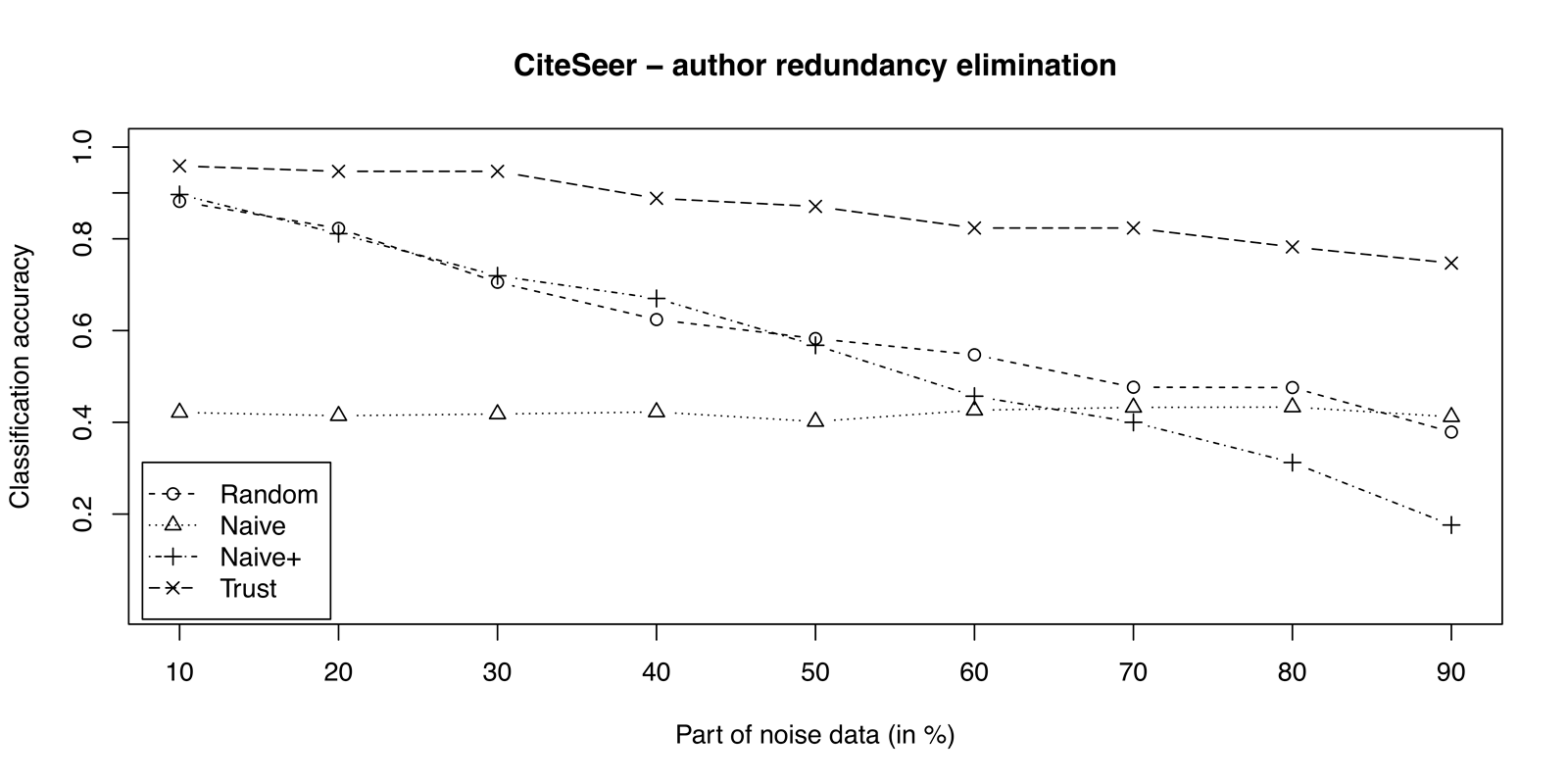}

}

\caption{Redundancy elimination on CiteSeer dataset for $34$ random clusters (as a result of ER), containing author names.}\label{fig:reCS}
\end{figure}

Author name attribute redundancy elimination results are presented on
Figure \ref{fig:reCS} and in Table \ref{tab:csRE}. The trust measure
achieves better results comparing to others. It is expected for accuracy
to be inversely proportional to level of noise, but the classification
accuracy of the trust is above \(70\%\) even with \(90\%\) of noise in
data. As we see, the trust measure outperforms other approaches
throughout the test. The naive measure gives almost constant accuracy at
all times. Naive+ approach performs vey bad by increasing the number of
noise values. The reason it works better than naive at low noise levels
is that there are many similar or equal values in cluster, but at higher
levels, the majority of values are quite different. It's results are
similar to the random measure. Random approach results are expected,
maybe even too good with clusters of a lot of noise. When having no
knowledge of cluster values, performance of redundancy elimination would
equal to random approach.

\begin{table}

\caption{\label{tab:csRE}Redundancy elimination classification accuracy on $34$ random author name clusters from CiteSeer dataset. The trust algorithm was repeated $10$-times and others $100$-times.}
\centering
\begin{tabular}[t]{ccccccccc}
\toprule
\multicolumn{1}{c}{ } & \multicolumn{2}{c}{Random} & \multicolumn{2}{c}{Naive} & \multicolumn{2}{c}{Naive+} & \multicolumn{2}{c}{Trust} \\
\cmidrule(l{2pt}r{2pt}){2-3} \cmidrule(l{2pt}r{2pt}){4-5} \cmidrule(l{2pt}r{2pt}){6-7} \cmidrule(l{2pt}r{2pt}){8-9}
Noise & Mean & Std. d. & Mean & Std. d. & Mean & Std. d. & Mean & Std. d.\\
\midrule
90\% & 0.37 & 0.078 & 0.41 & 0.083 & 0.17 & 0.063 & 0.74 & 0.084\\
80\% & 0.47 & 0.074 & 0.43 & 0.067 & 0.31 & 0.074 & 0.78 & 0.067\\
70\% & 0.47 & 0.069 & 0.43 & 0.075 & 0.39 & 0.070 & 0.82 & 0.046\\
60\% & 0.54 & 0.064 & 0.42 & 0.082 & 0.45 & 0.076 & 0.82 & 0.036\\
50\% & 0.58 & 0.062 & 0.40 & 0.084 & 0.56 & 0.081 & 0.87 & 0.033\\
\addlinespace
40\% & 0.62 & 0.067 & 0.42 & 0.071 & 0.67 & 0.071 & 0.88 & 0.038\\
30\% & 0.70 & 0.062 & 0.41 & 0.083 & 0.71 & 0.075 & 0.94 & 0.032\\
20\% & 0.82 & 0.038 & 0.41 & 0.078 & 0.81 & 0.061 & 0.94 & 0.038\\
10\% & 0.88 & 0.038 & 0.42 & 0.063 & 0.89 & 0.050 & 0.95 & 0.044\\
\bottomrule
\end{tabular}
\end{table}

The experiment shows it may be easy to get useful redundancy eliminator
for specific types of values, but the solution remains to initialize
trust levels across the domain and update them continuously during
system's lifetime.

\subsection{Experiments summary}\label{experiments-summary}

We presented some experiments on the attribute, entity resolution and
redundancy elimination components of the proposed general
\emph{framework} for matching and merging (see Figure
\ref{fig:framework}).

As first, attribute resolution matches the datasets to the same semantic
representation (see section \ref{attr-res}). When datasets are not
appropriately matched, missing attribute pairs cannot even be compared
or wrong values are considered. Therefore, further matching strongly
depends on attribute resolution result.

Second, we showed entity resolution improves if additional semantic
similarity measure is used (see section \ref{er-ex}). Semantic
similarity is attribute type-specific and cannot be defined in general.
Thus, a number of metrics could be predefined and then selected for each
attribute type.

Third, input to redundancy elimination are clusters as a result from
entity resolution (see section \ref{re-ex}). For author names, we showed
the search engine results as a value of trust, can help us determine the
most appropriate value. Also, this component's results strongly depend
on matched clusters results as only one value within specific cluster
can be selected.

To summarize, best evaluation measuring interdependence between
components could be achieved only when having a dataset annotated with
all needed contexts we defined. The proposed framework can be employed
for general tasks, but would be outperformed by domain-specific
applications.

\section{Discussion}\label{discussion}

Proposed framework for matching and merging represents a general and
complete solution, applicable in all diverse areas of use. Introduction
of contexts allows a joint control over various dimensions of matching
and merging variability, providing for specific needs of each scenario.
Furthermore, data architecture combines simple (network) data with
semantically enriched data, which makes the proposition applicable for
any data source. Framework can thus be used as a general solution for
merging data from heterogeneous sources, and also merely for matching.

The fundamental difference between matching, including only attribute
and entity resolution, and merging, including also redundancy
elimination, is, besides the obvious, in the fact that merged data is
read-only. Since datasets, obtained after merging, do not necessarily
resemble the original datasets, the data cannot be altered thus the
changes would apply also in the original datasets. Alternative approach
is to merely match the given datasets and to merge them only on demand.
When altering matched data, user can change the original datasets (that
are in this phase still represented independently) or change the merged
dataset (that was previously demanded for), in which case he must also
provide an appropriate strategy, how the changes should be applied in
the original datasets.

Proposed algorithms employ network data, semantically enriched with
ontologies. With the advent of \emph{Semantic Web}, ontologies are
gaining importance mainly due to availability of formal ontology
languages. These standardization efforts promote several notable uses of
ontologies like assisting in communication between people, achieving
interoperability (communication) among heterogeneous software systems
and improving the design and quality of software systems. One of the
most prominent applications is in the domain of semantic
interoperability. While pure semantics concerns the study of meanings,
semantic elevation means to achieve semantic interoperability and can be
considered as a subset of information integration (including data
access, aggregation, correlation and transformation). Semantic elevation
of proposed matching and merging framework represents one major step
towards this end.

Use of trust-aware techniques and algorithms introduces several key
properties. Firstly, an adequate trust management provides means to deal
with uncertain or questionable data sources, by modeling trustworthiness
of each provided value appropriately. Secondly, algorithms jointly
optimize not only entity resolution or redundancy elimination of
provided datasets, but also the trustworthiness of the resulting
datasets. The latter can substantially increase the accuracy. Thirdly,
trustworthiness of data can be used also for security reasons, by seeing
trustworthy values as more secure. Optimizing the trustworthiness of
matching and merging thus also results in an efficient security
assurance.

Although, contexts are merely a way to guide the execution of some
algorithm, their definition is relatively different from that of any
simple parameter. The execution is controlled with mere definition of
the contexts, when in the case of parameters, it is controlled by
assigning different values. For instance, when default behavior is
desired, the parameters still need to be assigned, when in the case of
contexts, the algorithm is used as it is. For any general solution,
working with heterogeneous clients, such behavior can significantly
reduce the complexity.

As different contexts are used jointly throughout matching and merging
execution, they allow a collective control over various dimensions of
variability. Furthermore, each execution is controlled and also
characterized with the context it defines, which can be used to compare
and analyze different executions or matching and merging algorithms.

Last, we briefly discuss a possible disadvantage of the proposed
framework. As the framework represents a general solution, applicable in
all diverse domains, the performance of some domain-specific approach or
algorithm can still be superior. However, such approaches commonly
cannot be generalized and are thus inappropriate for practical (general)
use.

\section{Conclusion}\label{conclusion}

This paper advances previously published paper
\citep{subelj_merging_2011} which contains only theoretical view of the
proposed framework for data matching and merging. In this work we again
overview the whole framework with minor changes, but most importantly we
introduce different metrics implementation details and full framework
demonstration.

The proposed framework follows a three level architecture using
network-based data representation from data to semantic and lastly to
abstract level. Data on each level is always a superset of lower ones
due to inclusion of various context types, trust values or additional
metadata. We also identify three main context types -- user, data and
trust context type -- which are a formal representation of all possible
operations. One of the novelties is also trust management that is
available across all steps during the execution.

To support our framework proposal, we conduct experiments of three main
components -- attribute resolution, entity resolution and redundancy
elimination -- using trust and semantics. Like we theoretically
anticipated, results on five datasets show that semantic elevation and
proper trust management significantly improve overall results.

In further work we will additionally incorporate network analysis
techniques such as community detection \citep{subelj_community_2011} or
recent research on self-similar networks
\citep{blagus_self-similar_2012}, which finds network hierarchies with a
number of common properties that may also improve the results of
proposed approach. Furthermore, ontology-based information extraction
techniques will be employed into entity resolution algorithm to gain
more knowledge about non-atomic values.

\section*{Acknowledgement}\label{acknowledgement}
\addcontentsline{toc}{section}{Acknowledgement}

We would also like to thank co-authors of the previous paper
\citep{subelj_merging_2011} -- David Jelenc, Eva Zupančič, Denis Trček
and Marjan Krisper -- who have helped us with theoretical knowledge at
framework design stage.

The work has been supported by the Slovene Research Agency ARRS within
the research program P2-0359 and part financed by the European Union,
European Social Fund.


\begin{thebibliography}{}

\bibitem[Adamic and Adar, 2001]{adamic_friends_2001}
Adamic, L. and Adar, E. (2001).
\newblock Friends and neighbors on the {Web}.
\newblock {\em Social Networks}, 25:211--230.

\bibitem[Ananthakrishna et~al., 2002]{ananthakrishna_eliminating_2002}
Ananthakrishna, R., Chaudhuri, S., and Ganti, V. (2002).
\newblock Eliminating fuzzy duplicates in data warehouses.
\newblock In {\em Proceedings of the {International} {Conference} on {Very}
  {Large} {Data} {Bases}}, pages 586--597.

\bibitem[Bengtson and Roth, 2008]{bengtson_understanding_2008}
Bengtson, E. and Roth, D. (2008).
\newblock Understanding the value of features for coreference resolution.
\newblock In {\em Proceedings of the {Conference} on {Empirical} {Methods} in
  {Natural} {Language} {Processing}}, pages 294--303. Association for
  Computational Linguistics.

\bibitem[Bhattacharya and Getoor, 2004]{bhattacharya_iterative_2004}
Bhattacharya, I. and Getoor, L. (2004).
\newblock Iterative record linkage for cleaning and integration.
\newblock In {\em Proceedings of the {ACM} {SIGKDD} {Workshop} on {Research}
  {Issues} in {Data} {Mining} and {Knowledge} {Discovery}}, pages 11--18.

\bibitem[Bhattacharya and Getoor, 2007]{bhattacharya_collective_2007}
Bhattacharya, I. and Getoor, L. (2007).
\newblock Collective entity resolution in relational data.
\newblock {\em ACM Transactions on Knowledge Discovery from Data}, 1(1):5.

\bibitem[Blagus et~al., 2012]{blagus_self-similar_2012}
Blagus, N., Šubelj, L., and Bajec, M. (2012).
\newblock Self-similar scaling of density in complex real-world networks.
\newblock {\em Physica A - Statistical mechanics and its applications},
  391(8):2794--2802.

\bibitem[{Blaze Software}, 1999]{blaze_software_blaze_1999}
{Blaze Software} (1999).
\newblock Blaze {Advisor}.
\newblock Technical {White} {Paper}, version 2.5.

\bibitem[Castano et~al., 2006]{castano_matching_2006}
Castano, S., Ferrara, A., and Montanelli, S. (2006).
\newblock Matching ontologies in open networked systems: {Techniques} and
  applications.
\newblock {\em Journal on Data Semantics}, pages 25--63.

\bibitem[Castano et~al., 2009]{castano_icoord_2009}
Castano, S., Ferrara, A., and Montanelli, S. (2009).
\newblock The {iCoord} knowledge model for {P}2p semantic coordination.
\newblock In {\em Proceedings of the {Conference} on {Italian} {Chapter} of
  {AIS}}.

\bibitem[Castano et~al., 2010]{castano_dealing_2010}
Castano, S., Ferrara, A., and Montanelli, S. (2010).
\newblock Dealing with matching variability of semantic {Web} data using
  contexts.
\newblock In {\em Proceedings of the {International} {Conference} on {Advanced}
  {Information} {Systems} {Engineering}}.

\bibitem[Chakrabarti et~al., 1998]{chakrabarti_automatic_1998}
Chakrabarti, S., Dom, B., Raghavan, P., Rajagopalan, S., Gibson, D., and
  Keinberg, J. (1998).
\newblock Automatic resource compilation by analyzing hyperlink structure and
  associated text.
\newblock {\em Proceedings of the International World Wide Web Conference},
  pages 65--74.

\bibitem[Cohen, 2000]{cohen_data_2000}
Cohen, W.~W. (2000).
\newblock Data integration using similarity joins and a word-based information
  representation language.
\newblock {\em ACM Transactions on Information Systems}, 18(3):288--321.

\bibitem[Cohen et~al., 2003]{cohen_comparison_2003}
Cohen, W.~W., Ravikumar, P., and Fienberg, S.~E. (2003).
\newblock A comparison of string distance metrics for name-matching tasks.
\newblock In {\em Proceedings of the {IJCAI} {Workshop} on {Information}
  {Integration} on the {Web}}, pages 73--78.

\bibitem[Domingos and Richardson, 2001]{domingos_mining_2001}
Domingos, P. and Richardson, M. (2001).
\newblock Mining the network value of customers.
\newblock In {\em Proceedings of the {International} {Conference} on
  {Knowledge} {Discovery} and {Data} {Mining}}, pages 57--66.

\bibitem[Dong et~al., 2005]{dong_reference_2005}
Dong, X., Halevy, A., and Madhavan, J. (2005).
\newblock Reference reconciliation in complex information spaces.
\newblock In {\em Proceedings of the {ACM} {SIGMOD} {International}
  {Conference} on {Management} of {Data}}, pages 85--96.

\bibitem[Euzenat and Shvaiko, 2007]{euzenat_ontology_2007}
Euzenat, J. and Shvaiko, P. (2007).
\newblock {\em Ontology matching}.
\newblock Springer-Verlag.

\bibitem[Gruber, 1993]{gruber_translation_1993}
Gruber, T.~R. (1993).
\newblock A {Translation} {Approach} to {Portable} {Ontology} {Specifications}.
\newblock {\em Knowledge Acquisition}, 5(2):199--220.

\bibitem[Hernandez and Stolfo, 1995]{hernandez_merge/purge_1995}
Hernandez, M. and Stolfo, S. (1995).
\newblock The merge/purge problem for large databases.
\newblock {\em Proceedings of the ACM SIGMOD International Conference on
  Management of Data}, pages 127--138.

\bibitem[Horrocks and Sattler, 2001]{horrocks_ontology_2001}
Horrocks, I. and Sattler, U. (2001).
\newblock Ontology reasoning in the {SHOQ} ({D}) description logic.
\newblock In {\em {IJCAI}}, volume~1, pages 199--204.

\bibitem[Jaro, 1989]{jaro_advances_1989}
Jaro, M.~A. (1989).
\newblock Advances in record linking methodolg as applied to the 1985 census of
  {Tampa} {Florida}.
\newblock {\em Journal of the American Statistical Society}, 84(406):414--420.

\bibitem[Kalashnikov and Mehrotra, 2006]{kalashnikov_domain-independent_2006}
Kalashnikov, D. and Mehrotra, S. (2006).
\newblock Domain-independent data cleaning via anlysis of entity-relationship
  graph.
\newblock {\em ACM Transactions on Database Systems}, 31(2):716--767.

\bibitem[Kautz et~al., 1997]{kautz_referral_1997}
Kautz, H., Selman, B., and Shah, M. (1997).
\newblock Referral {Web}: combining social networks and collaborative
  filtering.
\newblock {\em Communications of the ACM}, 40(3):63--65.

\bibitem[Kleinberg, 1999]{kleinberg_authoritative_1999}
Kleinberg, J.~M. (1999).
\newblock Authoritative sources in a hyperlinked environment.
\newblock {\em Journal of the ACM}, 46(5):604--632.

\bibitem[Köpcke et~al., 2010]{kopcke_evaluation_2010}
Köpcke, H., Thor, A., and Rahm, E. (2010).
\newblock Evaluation of entity resolution approaches on real-world match
  problems.
\newblock {\em Proceedings of the VLDB Endowment}, 3(1-2):484--493.

\bibitem[Lapouchnian and Mylopoulos, 2009]{lapouchnian_modeling_2009}
Lapouchnian, A. and Mylopoulos, J. (2009).
\newblock Modeling domain variability in requirements engineering with
  contexts.
\newblock In {\em Proceedings of the {International} {Conference} on
  {Conceptual} {Modeling}}, pages 115--130, Gramado, Brazil. Springer-Verlag.

\bibitem[Lavbič et~al., 2010]{lavbic_ontology-based_2010}
Lavbič, D., Vasilecas, O., and Rupnik, R. (2010).
\newblock Ontology-based multi-agent system to support business users and
  management.
\newblock {\em Technological and Economic Development of Economy},
  16(2):327--347.
\newblock https://sandbox.lavbic.net/publications/2010/TEDE.

\bibitem[Lee et~al., 2011]{lee_stanfords_2011}
Lee, H., Peirsman, Y., Chang, A., Chambers, N., Surdeanu, M., and Jurafsky, D.
  (2011).
\newblock Stanford's multi-pass sieve coreference resolution system at the
  {CoNLL}-2011 shared task.
\newblock In {\em Proceedings of the fifteenth conference on computational
  natural language learning: {Shared} task}, pages 28--34. Association for
  Computational Linguistics.

\bibitem[Lenzerini, 2002]{lenzerini_data_2002}
Lenzerini, M. (2002).
\newblock Data integration: {A} theoretical perspective.
\newblock In {\em Proceedings of the {ACM} {SIGMOD} {Symposium} on {Principles}
  of {Database} {Systems}}, pages 233--246.

\bibitem[Levenshtein, 1966]{levenshtein_binary_1966}
Levenshtein, V. (1966).
\newblock Binary codes capable of correcting deletions, insertions, and
  reversals.
\newblock {\em Soviet Physics Doklady}, 10(8):707--710.

\bibitem[Miller, 1995]{miller_wordnet:_1995}
Miller, G.~A. (1995).
\newblock {WordNet}: a lexical database for {English}.
\newblock {\em Communications of the ACM}, 38(11):39--41.

\bibitem[Monge and Elkan, 1996]{monge_field_1996}
Monge, A. and Elkan, C. (1996).
\newblock The field matching problem: {Algorithms} and applications.
\newblock {\em Proceedings of the International Conference on Knowledge
  Discovery and Data Mining}, pages 267--270.

\bibitem[Moreau et~al., 2008]{moreau_robust_2008}
Moreau, E., Yvon, F., and CappÈ, O. (2008).
\newblock Robust similarity measures for named entities matching.
\newblock In {\em Proceedings of the {International} {Conference} on
  {Computational} {Linguistics}}, pages 593--600.

\bibitem[Nagy et~al., 2008]{nagy_managing_2008}
Nagy, M., Vargas-Vera, M., and Motta, E. (2008).
\newblock Managing conflicting beliefs with fuzzy trust on the {Semantic}
  {Web}.
\newblock In {\em Proceedings of the {Mexican} {International} {Conference} on
  {Advances} in {Artificial} {Intelligence}}, pages 827--837.

\bibitem[Newman, 2010]{newman_networks:_2010}
Newman, M. (2010).
\newblock {\em Networks: an introduction}.
\newblock Oxford University Press, Oxford.

\bibitem[Ng, 2008]{ng_unsupervised_2008}
Ng, V. (2008).
\newblock Unsupervised models for coreference resolution.
\newblock In {\em Proceedings of the {Conference} on {Empirical} {Methods} in
  {Natural} {Language} {Processing}}, pages 640--649. Association for
  Computational Linguistics.

\bibitem[Rahm and Bernstein, 2001]{rahm_survey_2001}
Rahm, E. and Bernstein, P.~A. (2001).
\newblock A survey of approaches to automatic schema matching.
\newblock {\em Journal on Very Large Data Bases}, 10(4):334--350.

\bibitem[Resnick et~al., 1994]{resnick_grouplens:_1994}
Resnick, P., Iacovou, N., Suchak, M., Bergstrom, P., and Riedl, J. (1994).
\newblock {GroupLens}: {An} open architecture for collaborative filtering of
  netnews.
\newblock In {\em Proceedings of {ACM} {Conference} on {Computer} {Supported}
  {Cooperative} {Work}}, pages 175--186.

\bibitem[Richardson et~al., 2003]{richardson_trust_2003}
Richardson, M., Agrawal, R., and Domingos, P. (2003).
\newblock Trust management for the semantic {Web}.
\newblock In {\em Proceedings of the {International} {Semantic} {Web}
  {Conference}}, pages 351--368.

\bibitem[Thor and Rahm, 2007]{thor_moma_2007}
Thor, A. and Rahm, E. (2007).
\newblock Moma - a mapping-based object matching system.
\newblock In {\em Proceedings of the 3rd {Biennial} {Conference} on
  {Innovative} {Data} {Systems} {Research}}.

\bibitem[Trček, 2009]{trcek_formal_2009}
Trček, D. (2009).
\newblock A formal apparatus for modeling trust in computing environments.
\newblock {\em Mathematical and Computer Modelling}, 49(1-2):226--233.

\bibitem[Whang et~al., 2010]{whang_pay-as-you-go_2010}
Whang, S., Marmaros, D., and Garcia-Molina, H. (2010).
\newblock Pay-as-you-go {ER}.
\newblock Technical report.

\bibitem[Whang and Garcia-Molina, 2013]{whang_disinformation_2013}
Whang, S.~E. and Garcia-Molina, H. (2013).
\newblock Disinformation techniques for entity resolution.
\newblock In {\em Proceedings of the 22nd {ACM} international conference on
  {Information} \& {Knowledge} {Management}}, pages 715--720. ACM.

\bibitem[Winkler, 1990]{winkler_string_1990}
Winkler, W.~E. (1990).
\newblock String comparator metrics and enhanced decision rules in the
  {Fellegi}-{Sunter} model of record linkage.
\newblock In {\em Proceedings of the {Section} on {Survey} {Research}
  {Methods}}, pages 354--359.

\bibitem[Štajner and Mladenić, 2009]{stajner_entity_2009}
Štajner, T. and Mladenić, D. (2009).
\newblock Entity resolution in texts using statistical learning and ontologies.
\newblock In {\em Asian {Semantic} {Web} {Conference}}, pages 91--104.
  Springer.

\bibitem[Šubelj and Bajec, 2011a]{subelj_community_2011}
Šubelj, L. and Bajec, M. (2011a).
\newblock Community structure of complex software systems: {Analysis} and
  applications.
\newblock {\em Physcia A - Statistical mechanics and its applications},
  390(16):2968--2975.

\bibitem[Šubelj and Bajec, 2011b]{subelj_robust_2011}
Šubelj, L. and Bajec, M. (2011b).
\newblock Robust network community detection using balanced propagation.
\newblock {\em European Physical Journal B}, 81(3):353--362.

\bibitem[Šubelj et~al., 2011]{subelj_merging_2011}
Šubelj, L., Jelenc, D., Zupančič, E., Lavbič, D., Trček, D., Krisper, M.,
  and Bajec, M. (2011).
\newblock Merging data sources based on semantics, contexts and trust.
\newblock {\em IPSI BgD Transactions on Internet Research}, 7(1):18--30.

\end{thebibliography}
\end{document}